\documentclass[journal,comsoc,10pt]{IEEEtran}

\usepackage[T1]{fontenc}
\usepackage{cite}
\usepackage{amssymb,amsfonts,amsmath}
\usepackage{algorithmic}
\usepackage{graphicx}
\usepackage{multicol}
\usepackage{pgfplots}
\usepackage{subfigure}
\usepackage{caption}
\usepackage{tikz}
\usepackage{mathtools}
\usepackage{mwe}
\usepackage{textcomp}
\usepackage{xcolor}
\usepackage{balance}
\usepackage{multirow}
\usepackage[flushleft]{threeparttable}
\usepackage{booktabs} 
\def\BibTeX{{\rm B\kern-.05em{\sc i\kern-.025em b}\kern-.08em
    T\kern-.1667em\lower.7ex\hbox{E}\kern-.125emX}}
\usetikzlibrary{shapes.geometric, arrows}
\pgfplotsset{compat=newest} 
\pgfplotsset{plot coordinates/math parser=false} 
\newlength\figureheight 
\newlength\figurewidth
\graphicspath{ {Images/} }
\usepackage{balance}
\usepackage{enumitem}
\usepackage{amsmath}
\usepackage{bm}
\usepackage{url}
\usepackage{tabularx}

\newlist{abbrv}{itemize}{1}
\setlist[abbrv,1]{label=,labelwidth=1in,align=parleft,itemsep=0.1\baselineskip,leftmargin=!}

\begin{document}

\title{Pre-print: Radio Identity Verification-based IoT Security Using RF-DNA Fingerprints and SVM}

\author{Donald~Reising, 
        Joseph~Cancelleri, 
        T. Daniel Loveless, 
				Farah~Kandah, 
        and~Anthony~Skjellum
\thanks{D. Reising, J. Cancelleri, and T. Loveless are with the Electrical Engineering Department, University of Tennessee at Chattanooga, Chattanooga,
TN, 37403 USA e-mail: \{donald-reising,~gsy536,daniel-loveless\}@utc.edu.}
\thanks{F. Kandah is with the Computer Science Department, University of Tennessee at Chattanooga, Chattanooga,
TN, 37403 USA e-mail: farah-kandah@utc.edu.}
\thanks{A. Skjellum is the Director of the SimCenter, University of Tennessee at Chattanooga, Chattanooga,
TN, 37403 USA e-mail: anthony-skjellum@utc.edu.}
\thanks{Manuscript received THE DATE; revised SOME DATE.}}

\markboth{IEEE Internet of Things Journal,~Vol.~XX, No.~X, May~2020}%
{Reising \MakeLowercase{\textit{et al.}}: Bare Demo of IEEEtran.cls for IEEE Communications Society Journals}

\maketitle

\begin{abstract}
It is estimated that the number of IoT devices will reach 75 billion in the next five years. Most of those currently, and to be deployed, lack sufficient security to protect themselves and their networks from attack by malicious IoT devices that masquerade as authorized devices to circumvent digital authentication approaches. This work presents a PHY layer IoT authentication approach capable of addressing this critical security need through the use of feature reduced Radio Frequency-Distinct Native Attributes (RF-DNA) fingerprints and Support Vector Machines (SVM). This work successfully demonstrates 100\%: (i) authorized ID verification across three trials of six randomly chosen radios at signal-to-noise ratios greater than or equal to 6~{dB}, and (ii) rejection of all rogue radio ID spoofing attacks at signal-to-noise ratios greater than or equal to 3~{dB} using RF-DNA fingerprints whose features are selected using the Relief-F algorithm.
\end{abstract}

\begin{IEEEkeywords}
IoT Security, RF Fingerprinting, Support Vector Machines, Feature Selection, Radio ID Authentication.
\end{IEEEkeywords}

\IEEEpeerreviewmaketitle
\section{Introduction}\label{sec:introduction}%
\IEEEPARstart{T}{he} Internet of Things (IoT) is a collection of semi-autonomous, Internet connected devices comprised of inexpensive computing, networking, sensing, and actuation capabilities for sensing and acting within the physical world~\cite{DoD_IoT_2016}. The number of deployed IoT devices continues to explode annually and is estimated to reach roughly 75 billion by 2025~\cite{Gartner_2015,Juniper_2016,Statista_IoT_2019}. Alarmingly 70\% of all IoT devices do not employ encryption due to: (i) limited on-board computational capability, (ii) being too expensive for the manufacturer to implement, and (iii) scalability issues associated with implementation and management~\cite{Rawlinson_2014,Ray_CIC_2019}. This makes them, and the associated infrastructure, susceptible to attack by devices that are incorrectly authenticated as legitimate, when in fact they are not, due their use of compromised digital credentials that are transmitted in clear text. There is a critical need for an IoT security approach capable of defeating attacks in which illegitimate devices digitally masquerade as authorized IoT devices to circumvent digital authentication approaches. The need is exacerbated as bad actors exploit this weakness to conduct attacks against IoT infrastructure \cite{Larsen_CNN_2017,Wright_book_2015,Stanislav_Rapid_2015,Wright_Killer_2019,Simon_2016,Shipley_DEFCON_2014,Shipley_GitHub,Krebs_Mirai_2017}. \\
\indent Recently, the physical (PHY) layer approach known as Specific Emitter Identification (SEI) has been put forward as a solution capable of addressing this critical IoT need \cite{Talbot_CandS_2017,Sa_Access_2019}. One specific SEI implementation, known as Radio Frequency (RF) fingerprinting, facilitates radio discrimination by exploiting the unintentional `coloration' that is inherently imparted upon a radio's waveform during its generation and transmission \cite{ToonsKins95,UretenIEE99,DudczykSEI2,Jeffery_MobiCom_2007,JanaMobi08,BrikMobi08,Suski_IJESDF_2008,DanevIPSN09,Klein_ICC_2009,Liu_SEI_2009,Liu_SEI_2011,Kennedy_2010,Reising_IJESDF_2010,Reising_Dissertation,Williams_NSS_2010,Takahashi_CompApps_2010,TekbasIEE,Ellis_RadioSci,Soli_IEEE,CanadaSEI,azzouz,Wheeler_ICNC_2017,JafariMILCOM2018,Pan_2019,KoseAccess2019,Fadul_WCNC_2019,Tian_IOTJournal_2019,Wilson_GLOBECOM_2019,Kroon_AICompSci_2010,Cobb_IFS_11,Dubendorfer_MILCOM_2012,Rehman_JoCompSysSci_2014,Reising_InfoSec_2015,Patel_TransOnReli_2015,Baldini_Sciences_2018,Merchant_JSTSP_2018,AndrewsWiSec2019,KandahiThings2019}. The cause of this radio specific coloration is assigned to the inherent and distinct characteristics as well as interactions associated with the devices and components that make up a radio's RF front-end. The fact that RF fingerprint features are inherent and unique to a given radio makes them difficult to imitate; thus, making security approaches based upon them very difficult to bypass \cite{Wang_CommsMag_2016,Xu_CommsTuts_2016}.\\
\indent The majority of RF fingerprinting work has focused on radio \emph{classification (a.k.a., identification)}~\cite{ToonsKins95,UretenIEE99,DudczykSEI2,Jeffery_MobiCom_2007,JanaMobi08,BrikMobi08,Suski_IJESDF_2008,DanevIPSN09,Klein_ICC_2009,Liu_SEI_2009,Liu_SEI_2011,Kennedy_2010,Reising_IJESDF_2010,Reising_Dissertation,Williams_NSS_2010,Takahashi_CompApps_2010,TekbasIEE,Ellis_RadioSci,Soli_IEEE,CanadaSEI,azzouz,Wheeler_ICNC_2017,JafariMILCOM2018,Merchant_JSTSP_2018,Pan_2019,KoseAccess2019,Fadul_WCNC_2019,Tian_IOTJournal_2019,Wilson_GLOBECOM_2019}. In classification, an \emph{unknown} radio's identity is determined through the comparison of its RF fingerprint(s) with each of the stored, reference models that represent the \emph{authorized/known} radios; thus, representing a ``one-to-many'' comparison. The radio identity associated with the reference model that results in the ``best'' match is said to be the originator of the RF fingerprint(s). The flaw in classification is that class assignment, in this case the radio assigned as the RF fingerprint originator, is made no matter how poor this ``best'' match is. This flaw can result in the granting of network access to \emph{rogue} radios. As defined in \cite{Reising_InfoSec_2015}, a rogue radio is one that intentionally falsifies its digital credentials, e.g., compromised passwords or encryption keys, spoofed MAC addresses or equivalent digital identities, to match that of an authorized radio to bypass digital security mechanisms.\\
\indent The flaw of radio classification has led to the proposal of a ``one-to-one'' comparison known as radio identity (ID) \emph{verification (a.k.a., authentication)}~\cite{Kroon_AICompSci_2010,Cobb_IFS_11,Dubendorfer_MILCOM_2012,Rehman_JoCompSysSci_2014,Reising_Dissertation,Reising_InfoSec_2015,Talbot_CandS_2017,Baldini_Sciences_2018,Merchant_JSTSP_2018,AndrewsWiSec2019,KandahiThings2019}. In radio ID verification the RF fingerprint(s) of the unknown radio are compared \emph{only} to the stored reference model associated with the presented digital ID \cite{Reising_InfoSec_2015}. Thus, the unknown radio's ID is either verified as that of an authorized radio or rejected as a rogue. In the presence of rogue radios, radio ID verification results in four possible outcomes, which were introduced in \cite{Reising_InfoSec_2015} and are presented in Table~\ref{tbl:VerifyErrors}. \\
\begin{table}[!b]
\vspace*{-0.05in}
\renewcommand{\arraystretch}{1.4}
  \caption{Identity Verification Outcomes \cite{Reising_InfoSec_2015}.}
   \label{tbl:VerifyErrors}
  \centering
  \begin{tabular}{ccc}
  \hline
  {} & \multicolumn{2}{c}{System Declaration}\\
  \cline{2-3}
  Actual ID & Authorized & Rogue \\
  \hline
  Authorized & True Verification (TVR) & False Reject (FRR) \\
  Rogue & False Verification (FVR) & True Reject (TRR) \\
  \hline
  \end{tabular}
  \vspace{-4mm}
\end{table}
\indent In this work, we present a radio ID verification-based IoT security approach using RF-Distinct Native Attributes (RF-DNA) fingerprints and Support Vector Machines (SVM). Our work differs from prior RF fingerprinting-based ID verification approaches in that feature selection is: (i) assessed using eight different techniques and (ii) driven by the authorized radio whose ID is to be verified. The latter is based upon the one-to-one nature of radio ID verification, which allows the RF-DNA fingerprints to differ in composition and number of features from one authorized radio to another. The key is that prior to feature selection, the RF-DNA fingerprints of every radio, i.e., authorized and rogue, are generated to be comprised of the same set of features. Lastly, `best' SVM model selection is performed using a novel approach that does not require any knowledge of the rogue radios' RF-DNA fingerprints.\\
\indent The remainder of this paper is structured as follows. First, a description of the adopted threat model is presented in Section~\ref{sec:threat_model}. Section~\ref{sec:related_work} provides a summary of prior RF fingerprint-based ID verification publications. The contributions of this work to the RF fingerprint-based ID verification state-of-the-art is presented in Section~\ref{sec:contribution}. Section~\ref{sec:methodology} outlines each stage within the presented ID verification process: signal collection, detection, and post-processing (Sect.~\ref{sec:SOI}), RF-DNA fingerprint generation (Sect.~\ref{sec:RFF_Gen}), SVM (Sect.~\ref{sec:SVM}), selection of the `best' SVM model (Sect.~\ref{sec:select_svm}), as well as the ID verification approach (Sect.~\ref{sec:id_approach}). The Methodology is followed by the ID verification and rogue rejection performance results and analysis in Section~\ref{sec:results}, which includes assessment of the developed approach using the eight feature selection approaches (Sect.~\ref{sec:results_feature_select}) and under degrading channel noise conditions (Sect.~\ref{sec:degrading_snr}). Lastly, Section~\ref{sec:conclusion} concludes the paper.
\section{Threat Model}\label{sec:threat_model}%
The threat model adopted in this work is informed by the prior RF-DNA fingerprinting work in \cite{Dubendorfer_MILCOM_2012,Patel_TransOnReli_2015,Reising_InfoSec_2015} as well as the threat models presented in \cite{Clancy_CrownCom_2008,Xie_TransMobi_2020}. In this work, the adversary leverages commercially available IoT and compute devices to conduct their attack upon an IoT infrastructure. It is assumed that the adversary has either the knowledge or access to simple software applications that enable them to modify the settings of their associated IoT device and compute devices necessary to conduct their attack \cite{Xie_TransMobi_2020}. The adversary is not an authorized user of the targeted IoT infrastructure; thus, they do not inherently have access to its wireless network(s) nor the individual IoT and support devices that form the infrastructure. Lastly, it is assumed that the IoT infrastructure communication links and hardware are not compromised.

\begin{figure}[!t]
  \centering
  \includegraphics[width=0.9\columnwidth]{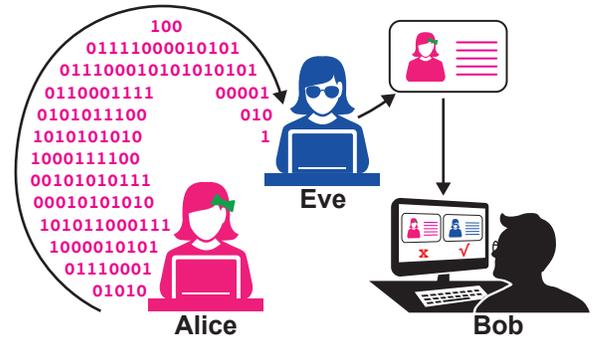}
    \caption{\underline{Threat Model}: Lacking encryption, the \emph{rogue} device `Eve' is able to trick `Bob', another authorized IoT device or IoT infrastructure monitor, that she is the authorized IoT device `Alice' by presenting Alice's digital ID credentials.}
    \vspace{-5mm}
    \label{fig:threat_model}
\end{figure}

Figure~\ref{fig:threat_model} provides an illustration of the particular IoT attack of interest within this work. The adversary, a.k.a., Eve, is able to gain access to the IoT infrastructure by exploiting the lack of encryption or easily circumvents an incorrectly implemented encryption process. Once encryption is bypassed, Eve is then authenticated by `Bob', an authorized IoT device or IoT infrastructure monitoring device, by presenting the compromised digital credentials, e.g., MAC address, International Mobile Subscriber Identity (IMSI), of another authorized IoT device, a.k.a., `Alice'. Following authentication, Eve is granted access to the IoT infrastructure and is able to: observe, spoof, inject, remove, and alter data in real-time as well as spoof the ID of other authorized IoT devices.
\section{Related Work}\label{sec:related_work}%
This section highlights the contributions and differences of previous RF fingerprint-based ID verification works.

In~\cite{Dubendorfer_MILCOM_2012}, RF-DNA fingerprint-based radio ID verification is performed using nine IEEE 802.15.4 ZigBee radios, seven authorized and two rogue, Fisher-based feature selection, and a probabilistic-based verification approach. Using a FVR$\leq$10\%, eleven out of fourteen rogue attacks, i.e., each rogue spoofs the ID of each authorized radio, were detected at a signal-to-noise  ratio  (SNR) of 10~{dB}. The goal is to achieve a FVR$\leq$10\% for all fourteen rogue radio attacks. In~\cite{Dubendorfer_MILCOM_2012}, the models used for radio authentication are developed for the purpose of {classification} and \emph{not} verification, which presents a potential reason for the poorer rogue radio rejection performance. \\
\indent A rogue radio rejection process based upon the $k$-Nearest Neighbor (KNN) classifier is presented in \cite{Rehman_JoCompSysSci_2014}. The presented approach achieves an accuracy from a low of 30\% up to 94\% at SNR$=$15~{dB}. KNN suffers from: (i) high outlier sensitivity, (ii) lacks kernel functions that aid in handling nonlinear data, (iii) requiring hyperparameters that are precisely fine tuned to achieve optimal results, and (iv) being poorly suited to unpredictable cases.

In \cite{Reising_InfoSec_2015}, RF-DNA fingerprints are used to verify the identity of Worldwide Interoperability for Microwave Access (WiMAX) radios in the presence of rogue radio attacks. Near-transient signals are collected from a total of eighteen WiMAX radios of the same manufacturer and model. The eighteen WiMAX radios are divided into three trials of six each. For a given trial, the remaining twelve radios serve as the rogues for that trial, which results in a total of 72 rogue radio attacks per trial. For the three trials, the highest number of rejected rogue radio attacks are 67 of 72 at a TRR$\geq$90\% and SNR$=$21~{dB}. In an effort to reject all rogue radio attacks, a fourth trial was selected. This fourth trial consisted of eight radios and successfully rejected all rogue radio attacks at SNR$=$21~{dB}. As in~\cite{Dubendorfer_MILCOM_2012}, the models and RF-DNA fingerprint feature selection was performed to maximize \emph{classification} performance. 

The work in \cite{Patel_TransOnReli_2015} performs radio ID verification of ZigBee radios using RF-DNA fingerprints and verification approaches using test metrics from three classifiers: Random Forest (RndF), Multiple Discriminant Analysis/Maximum Likelihood (MDA/ML), and Generalized Relevance Learning Vector Quantization-Improved (GRLVQI). The RndF and MDA/ML-based approaches use posterior probability, while the GRLVQI-based verification approach uses the angle-distance metric from \cite{Reising_InfoSec_2015}. The authors in \cite{Patel_TransOnReli_2015} perform feature selection using the: (i) feature importance ranking built-in to the RndF classifier, (ii) Kolmogorov-Smirnov test, and (iii) Dimensional Reduction Analysis (DRA) first presented in \cite{Reising_InfoSec_2015}.  Radio ID verification and rogue rejection is conducted using a total of thirteen ZigBee radios in which four are designated as authorized radios and the remaining nine assigned to serve as rogue radios; thus, representing a total of \emph{thirty six} rogue attacks, i.e., each rogue spoofs the digital ID of each authorized radio. For a TRR$\geq$90\%, 31 of 36 ($\sim$86\%) rogue radio spoofing attacks are rejected using DRA selected features and RndF-based radio ID verification at an SNR$=$12~{dB}. As with the RF-DNA fingerprint-based work in \cite{Dubendorfer_MILCOM_2012,Reising_InfoSec_2015}, the work in \cite{Patel_TransOnReli_2015} performs radio ID verification and rogue rejection using selected features and authorized radio models originally developed for the purpose of performing ``one-to-many'' \emph{classification}, which may have contributed to the poorer rogue radio rejection performance.

In \cite{Talbot_CandS_2017}, rogue attacks on IoT and commercial home automation systems are detected through the use of Slope-Based Frequency Shift Keyed (SB-FSK) fingerprints and a MDA/ML-based ID verification process. The attacks are carried out by like-model Insteon Switch (IS) devices as well as YARD Stick One Software Defined Radios (SDRs). At an SNR$=$15~{dB}, a total of 25 IS and 36 YARD Stick One SDR rogue attacks are successfully rejected for a TRR$=$95\% (FVR$=$5\%) and TRR$=$100\% (FVR$=$0\%), respectively. As in \cite{Dubendorfer_MILCOM_2012,Reising_InfoSec_2015,Patel_TransOnReli_2015}, the work in \cite{Talbot_CandS_2017} relies upon features and machine learning models that are optimized for classification not ID verification performance.

SVM-based ID verification, using RF fingerprints drawn from the preambles of the Random Access Channel (RACH) waveforms, of five 3GPP UMTS mobile radios is presented in \cite{Kroon_AICompSci_2010}. For ID verification, SVM is employed using both a single and ensemble-based approach. For ensemble-based ID verification three configurations are used: tiered, weighted tiered, and double weighted tiered. All of the ensemble-based approaches are implemented using a two class SVM in which each class represents a single radio. This approach requires the development of an SVM model for all possible paired combinations. Given radios A, B, and C, SVM models are developed for pairs: (A and B), (B and C), and (A and C). For the case when a radio presents the digital credentials of A, then its corresponding RF fingerprints are compared with two SVM models: (A and B) and (A and C). Verification of the radio's ID is dependent upon which ensemble approach is selected, but all three consider the class decision returned by each SVM model. \\
\indent In \cite{Merchant_JSTSP_2018}, ID verification is presented using Convolutional Neural Networks (CNN) and RF fingerprints learned from the waveforms of seven IEEE 802.15.4 ZigBee compliant radios. Similar to the work in \cite{KandahiThings2019} as well as this work, the work in \cite{Merchant_JSTSP_2018} performs ID verification using a two class model in which one class represents the authorized radio, whose ID is to be verified, and the other represents all remaining known radios, i.e., the not A case. Therefore, a total of seven CNNs are trained to perform ID verification. The use of CNN is more computationally and temporally complex than SVM. An average TVR$=$92.7\% is achieved at an SNR$=$10~{dB} while individual ID verification results are not presented. \\
\indent The work in \cite{KandahiThings2019} presents IoT device authentication implemented via radio ID verification, as described in Sect.~\ref{sec:introduction}, using RF-DNA fingerprints extracted from the near-transient responses of eighteen IEEE 802.16e WiMAX radios, the DRA feature selection approach from \cite{Reising_InfoSec_2015}, and an SVM classifier. Six WiMAX radios were designated as authorized radios and the remaining twelve used as rogue radios. An SVM model is developed for each of the six authorized radios using the RF-DNA fingerprints for all authorized WiMAX radios, where class one represents the radio whose digital ID needs verification, and class two represents the remaining five authorized WiMAX radios. The presented approach performed well in verifying the ID of five out of the six authorized radios while rejecting the rogue radios at a rate of 91\% or better SNR$\geq$9~{dB}. However, for the sixth authorized radio roughly 88\% of the rogue radios' transmissions were verified, i.e., FVR$=$88\%, as having originated from this sixth radio at SNR$=$9~{dB}. The FVR did not improve with increased SNR, which suggests that the issue lies either with the DRA feature selection approach and/or the SVM classifier. In~\cite{KandahiThings2019}, the DRA features were not chosen for the purpose of verifying the ID of a given authorized radio in mind, but to optimize one-to-many \emph{classification} performance. \\
\indent A Mobile IoT (MIoT) device ID verification approach using RF fingerprints and Support Vector Data Description (SVDD) is presented in \cite{Tian_Sensors_2020}. The approach uses Principal Component Analysis (PCA) followed by Neighborhood Component Analysis (NCA) to generate the RF fingerprint features from the instantaneous amplitude envelope of the power-on transient signals collected from ten Motorola A12 radios. The presented approach achieves TVR$\in$[86, 96]\% across five authorized radio trials at an SNR$=$15~{dB}. Each trial is comprised of eight authorized radios to facilitate rogue radio rejection assessment using that trial's two remaining radios. The results in \cite{Tian_Sensors_2020} achieve FVR$\in$[6, 9]\% for the rogue radios across the five trials at SNR$=$15~{dB}. In \cite{Tian_Sensors_2020}: (i) the RF fingerprints are extracted from the signal's instantaneous amplitude that is negatively affected by noise, (ii) only a single feature selection approach is considered, (iii) the RF fingerprints of every radio are comprised of the same number of features, and (iv) TVR$\geq$90\% requires a SNR$\geq$20~{dB}.
\begin{table*}[!t]
\vspace*{-0.05in}
\renewcommand{\arraystretch}{1.4}
  \caption{List of Notations.}
  \label{tbl:notation}
  \centering
  \begin{tabular}{|c|m{.25\textwidth}||c|m{.25\textwidth}||c|m{.25\textwidth}|}
  \hline
    $N_{D}$ & Number of radios used in this work &
    $N_{B}$ & Number of collected signals per radio &
    $N_{O}$ & Order of the Butterworth filter \\
    $G_{mk}$ & Complex Gabor Transform Coefficients &
    $M$ & Total number of time shifts &
    $N_{\Delta}$ &  Step size between Gabor calculations \\
    $K_{G}$ & Total number of frequency shifts &
    $N_{P}$ & Total patches in the time-frequency plane &
    $N_{T}$ & Size of a patch in time \\
    $N_{F}$ & Size of a patch in frequency &
    $P_{tf}$ & A single patch in the time-frequency plane &
    $\sigma$ &  Standard deviation \\
    $\sigma^{2}$ & Variance &
    $\gamma$ & Skewness &
    $\kappa$ & Kurtosis \\
    $N_{f}$ & Number of RF-DNA fingerprint features &
    $\lambda$ & DRA feature relevance vector &
    $\mathbf{f}$ & A single RF-DNA fingerprint \\
    $\mathbf{F}$ & A set of RF-DNA fingerprints &
    $\mathbf{w}$ & A projection matrix &
    $c_{i}$ &  The $i^{\text{th}}$ SVM class \\
    $N_{t.i}$ & Training fingerprints of class $i$ &
    $N_{\tau}$ & Total number of training fingerprints &
    $\mathbf{\bar{F}}$ & A set of normalized RF-DNA fingerprints \\
    $e$ & Eigenvectors &
    $\lambda_{e}$ & Eigenvalues &
    $N_{r}$ & Number of retained fingerprint features \\
    $w_{r}$ & Weight assigned to feature $r$ &
    $\mathcal{B}_{r}$ & Bhattacharyya coefficient of feature $r$ &
    $N_{\mathcal{B}}$ & Number of histogram bins \\
    $\mathbf{\mathcal{F}}$ & Set of RF-DNA fingerprints with class labels &
    $\Upsilon$ & A kernel function &
    $\rho$ & Vector of Probability of Error values \\
    $\alpha$ & Average Correlation Coefficient values &
    $w_{\rho}$ & Probability of Error weight & 
    $w_{\alpha}$ & Average Correlation Coefficient weight \\
    $\mathbf{t}$ & Statistic produced by the $t$-test & 
    $N_{K}$ & Number of Relief-F nearest neighbors &
    $\tilde{\mathbf{f}}$ & Randomly chosen RF-DNA fingerprint \\
    $\mathbf{f}_{H}$ & Relief-F nearest hit fingerprint &
    $\mathbf{f}_{M}$ & Relief-F nearest miss fingerprint &
    $\beta$ & An SVM support vector \\
    \hline
  \end{tabular}
  \vspace{-4mm}
\end{table*}
\section{Motivation and Contributions}\label{sec:contribution}%
In comparison to our approach, prior publications present results generated using: (i) a single feature selection approach, (ii) a set of RF fingerprint features that remain fixed in both number and composition across all authorized and rogue radios, (iii) machine learning models that are developed to maximize classification and \emph{not} ID verification performance, and (iv) threshold-driven performance. The work presented herein provides the following contributions to the area of SEI-based IoT security:
\begin{enumerate}[leftmargin=*]
\item{Feature selection is performed with ID verification in mind; thus, the number and particular RF-DNA fingerprint features retained are allowed to differ from one authorized radio to another. 
The use of the same set of features, for every authorized radio, overlooks the specific features that make a given authorized radio unique, which impedes ID verification and rogue rejection performance as SNR degrades.}
\item{SVM model development is performed for the sole purpose of performing ID verification and rogue rejection for a given authorized radio. Therefore, the SVM is trained using a two class approach where: class 1 represents the authorized radio whose ID needs to be verified and class 2 represents all other radios both rogues and remaining authorized radios. During SVM training, class 2 is developed using the RF-DNA fingerprints of all remaining authorized radios, i.e., those authorized radios whose ID is \emph{not} being verified, to serve as a representative sample of possible rogue radios.}
\item{ID verification and rogue radio rejection performance is assessed using eight feature selection approaches.
}
\item{A novel approach for selecting the SVM model that is `best' suited to simultaneously maximizing authorized radio ID verification and rogue radio rejection performance without any knowledge of the rogue radios’ RF-DNA fingerprints.}
\item{ID verification and rogue radio rejection performance does \emph{not} require the setting of a threshold. Threshold-based approaches require a trade-off between the rate at which the authorized radios' IDs are verified and rogue radios are rejected. So, increasing the ID verification rate requires sacrificing rogue rejection performance and vice versa. This dependency between the verification and rejection rates leads to degraded performance at lower SNR values.}
\item{Comparative assessment is facilitated by adopting the TVR$\geq$90\% and FVR$\leq$10\% benchmarks used in \cite{Dubendorfer_MILCOM_2012,Reising_InfoSec_2015,Patel_TransOnReli_2015,Talbot_CandS_2017,KandahiThings2019}. The goal is for the developed RF fingerprint-based ID verification approach to satisfy both of these benchmarks at the lowest SNR possible.}
\end{enumerate}
These contributions result in an average TVR$=$97.8\% at an SNR$=$6~{dB} while rejecting all rogue radio attacks at an FVR$\leq$10\% for SNR$\geq$3~{dB}, which is unseen in published literature.
\section{Our Methodology}\label{sec:methodology}%
Our proposed ID verification approach is carried out through the use of six consecutive phases, which are: (i) signal collection, detection \& post-processing, (ii) RF-DNA fingerprint generation, (iii) feature selection, (iv) SVM model development, (v) SVM model selection for ID verification under rogue radio attacks, and (iv) the ID verification approach.
\subsection{Signal Collection, Detection, and Post-Processing\label{sec:SOI}}%
\begin{figure}[!t]
  \centering
  \includegraphics[width=0.9\columnwidth]{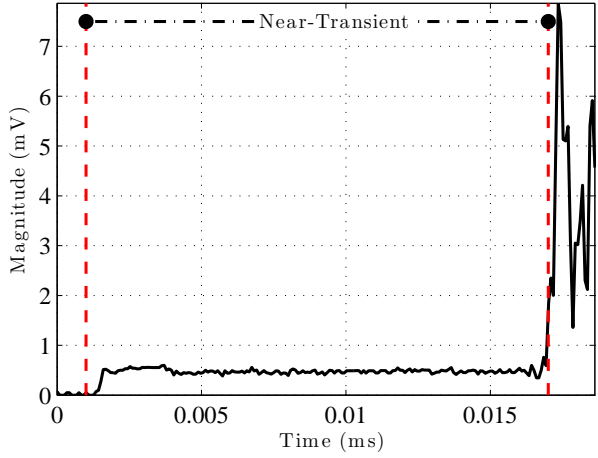}
    \caption{Representative ``near-transient'' response from a WiMAX MS radio's range-only transmission within the up-link sub-frame~\cite{Reising_InfoSec_2015}.}
    \vspace{-5mm}
    \label{fig:near_transient}
\end{figure}

This section provides a brief explanation of the process used to: (i) collect the signals from each of the eighteen WiMAX radios, (ii) separate individual transmissions from the overall collection record, and (iii) prepare the detected transmissions for RF-DNA fingerprint generation.\\
\indent RF-DNA fingerprints are drawn from the near-transient response present at the start of each range-only transmission generated by a WiMAX Mobile Subscriber (MS) radio within the up-link sub-frame. A representative illustration of this near-transient response is shown in Fig.~\ref{fig:near_transient}. Use of the WiMAX near-transient is due to the use of the on-hand data set in \cite{Reising_InfoSec_2015,KandahiThings2019}, which enables comparative analysis.

The near-transient responses of $N_{D}$$=$$18$ Alvarion BreezeMAX Extreme 5000 802.16e WiMAX MS radios are collected using an Agilent spectrum analyzer. 
The spectrum analyzer: (i) has an RF bandwidth of 36~{MHz}, (ii) operates over the range of frequencies from 20~{MHz} to 6~{GHz}, (iii) has a maximum sampling rate of 95~{mega-sample/s}, and a 12-bit analog-to-digital converter~\cite{Agilent}. Amplitude-based variance trajectory detection was used to select a total of $N_{B}$$=$$1,000$ near-transient responses for each of the WiMAX MS radios \cite{Klein_ICC_2009}. All detected near-transient responses are filtered using a $N_{O}$$=$$6^{\text{th}}$ order Butterworth filter and the In-phase and Quadrature (IQ) samples stored for RF-DNA fingerprint generation.

\subsection{RF-DNA Fingerprint Generation\label{sec:RFF_Gen}}%
In this section, the process used for generating an RF-DNA fingerprint, from a given collected signal, is explained. RF-DNA fingerprints are generated from the WiMAX near-transient response's time-frequency (TF) representation, which is calculated using the Discrete Gabor Transform (DGT) \cite{Bastiaans96}. Use of the DGT is due to its: (i) computational complexity being proportional to the sampling rate, (ii) robustness to degrading SNR when the calculation is \emph{over-sampled}, and (iii) demonstrated success in prior RF-DNA fingerprinting work \cite{Reising_InfoSec_2015,KandahiThings2019}, which facilitates comparative assessment. For all results presented in Sect.~\ref{sec:results}, the DGT is calculated by,
\begin{equation}
G_{mk} = 
\sum\limits_{n=1}^{MN_\Delta}{s(n+lMN_\Delta)\nu^{\ast}(n-mN_\Delta)exp^{-j2\pi kn/K_G},}
\label{eq:dgt}
\end{equation}
where $G_{mk}$ are Gabor coefficients, $s(n)$ is the near-transient response, $\nu(n)$ is a Gaussian analysis window, $m$$=$$1, 2, \dots, M$ for $M$ total shifts along the time dimension, and $k$$=$$0, 1, \dots, K_{G}-1$. Selection of $M$, $N_{\Delta}$, and $K_{G}$ must result in the modulus of $(MN_{\Delta})$ and $K_{G}$ returning zero. For consistency with \cite{Reising_InfoSec_2015,KandahiThings2019}, the DGT is calculated with $M$$=$$150$, $K_{G}$$=$$150$, and $N_{\Delta}$$=$$1$, which results in \emph{oversampling} due to  $K_{G}$$>$$N_{\Delta}$.

\begin{figure}[!t]
  \centering
  \includegraphics[width=0.9\columnwidth]{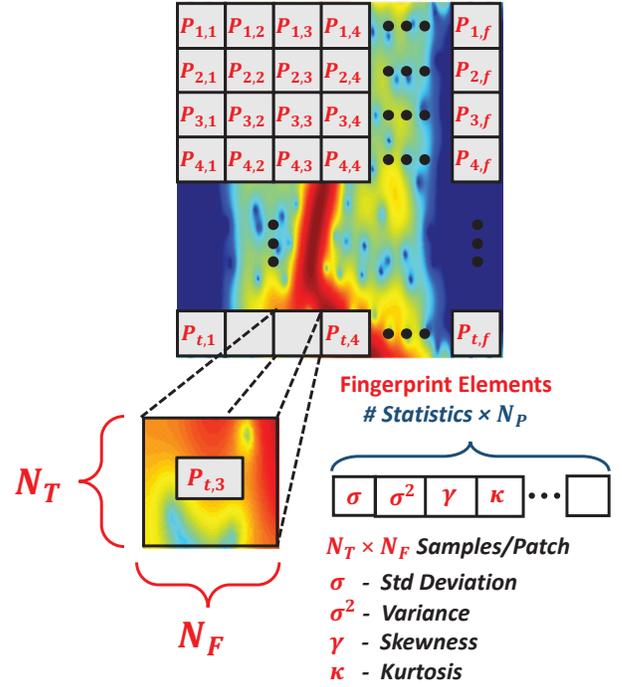}
    \caption{Illustration of Gabor-based RF-DNA fingerprint generation using $N_{T}$$\times$$N_{F}$ two-dimensional patches extracted from the centered, normalized magnitude-squared Gabor coefficients.}
    \vspace{-5mm}
    \label{fig:gabor_print_gent}
\end{figure}

The RF-DNA fingerprints are extracted from the normalized, magnitude-squared Gabor coefficients. The normalization ensures that all values, within the resulting TF representation, range between zero and one. Figure~\ref{fig:gabor_print_gent} shows the RF-DNA fingerprint generation process. 
The normalized TF representation is subdivided into $N_{P}$$=$$50$ patches. Each patch is comprised of $N_{T}$$=$$15$ by $N_{F}$$=$$10$ entries and is annotated as $P_{tf}$, where $t$ and $f$ denotes a particular patch's position within the TF response. Following selection of a particular patch, it is reshaped into a $N_{T}$$\times$$N_{F}$$=$$150$ length vector and the features: standard deviation $(\sigma)$, variance $(\sigma^{2})$, skewness $(\gamma)$, and kurtosis $(\kappa)$ calculated. This process is repeated for all patches and subsequent features are appended to the end of those calculated from previous patches. Additionally, these features are calculated for the entire normalized TF representation and added as the final four features of the RF-DNA fingerprint. Thus, each RF-DNA fingerprint is comprised of $N_{f}$$=$$204$ total features.

\subsection{Feature Selection Approaches}\label{sec:feat_methods}%
This work investigates eight feature selection techniques: DRA, Linear Discriminant Analysis (LDA), PCA, NCA, Probability Of Error plus Average Correlation Coefficient (POEACC), Bhattacharyya Coefficient (BC), $t$-test, and Relief-F \cite{Reising_Dissertation,dhsPC,YangJCP2012,MucciardiITC,Comaniciu_CVPR_2000,Derrick_TQMP_2016,Allwood_APStats_2008,Kononenko}. The features that comprise an RF-DNA fingerprint are the statistics, i.e., variance, standard deviation, skewness, kurtosis, calculated from the normalized, magnitude-squared Gabor response as described in Section~\ref{sec:RFF_Gen}. Feature selection is performed for each authorized radio with the goal of maximizing the TVR while simultaneously minimizing the FVR. This is facilitated by allowing the number of retained features $N_{r}$ as well as which features are selected to vary from one authorized radio's set of RF-DNA fingerprints to another. The number of retained features $N_{r}$ is chosen to be the fewest that result in a TVR$\geq$90\% and FVR$\leq$10\%. The remainder of this section explains the feature selection process for each of the eight selected approaches.
\subsubsection{Dimensional Reduction Analysis\label{sec:DRA}}%
DRA is a feature selection approach first introduced in \cite{Reising_Dissertation} and leveraged in \cite{Reising_InfoSec_2015,KandahiThings2019} to reduce the dimensionality of the RF-DNA fingerprints prior to radio ID verification. DRA selects the most relevant features based upon the values contained in the feature relevance vector, $\lambda$,  which is generated as part of the GRLQVI classifier's training phase. At a given SNR, the relevance vector is,
\begin{equation}
    \mathbf{\lambda}(\text{SNR}) = \left[\lambda_{1},\lambda_{2},\dots,\lambda_{N_{f}}\right],
\end{equation}
where $\lambda_{j}$$\in$$[0,1]$ indicates how much feature $j$ influences the classification decision. If $\lambda_{j}$$>$$\lambda_{k}$, then feature $j$ is more influential on the classification decision than that of feature $k$. For the presented results, feature selection is performed on a per SNR basis using the feature relevance vector corresponding to the GRLVQI classification model learned at that SNR.
\subsubsection{Linear Discriminant Analysis\label{sec:LDA}}%
LDA is a linear operation that projects the $N_{f}-$dimensional RF-DNA fingerprints onto a line that results in the greatest separation of the two classes of interest. In this work, the two classes are the set of RF-DNA fingerprints corresponding to: (i) the authorized radio ($c_{1}$), whose ID is to be verified, and (ii) all other authorized radios ($c_{2}$) that serve as representatives of all other WiMAX radios including rogues. The RF-DNA fingerprints are projected onto the line by,
\begin{equation}
    \mathbf{F}_{w} = \mathbf{w}^{T}\mathbf{F},
    \label{eq:fingerprint_projection}
\end{equation}
\begin{equation}
\resizebox{.91\columnwidth}{!}{$
    \mathbf{F} = \left[\mathbf{f}_{1}^{c_{1}},\mathbf{f}_{2}^{c_{1}}, \dots, \mathbf{f}_{N_{t,1}}^{c_{1}}, \dots, \mathbf{f}_{1}^{c_{2}},\mathbf{f}_{2}^{c_{2}}, \dots, \mathbf{f}_{N_{t,2}}^{c_{2}}\right]_{N_{\tau}\times N_{f}}^{T},$}
    \label{eq:training_fingerprint_set}
\end{equation}
where $\mathbf{f}$ is an $N_{f}-$dimensional RF-DNA fingerprint belonging to class $c_{1}$ or $c_{2}$, $N_{t,i}$ is the total number of RF-DNA fingerprints in the training set of class $i$, $N_{\tau}$$=$$(N_{t,1} + N_{t,2})$, and the superscript $T$ denotes transpose. The projection $\mathbf{w}$ is,
\begin{equation}
    \mathbf{w} = \mathbf{S}_{w}^{-1}(\mathbf{\mu}_{1} - \mathbf{\mu}_{2}),
    \label{eq:projection_matrix}
\end{equation}
where
\begin{equation}
    \mathbf{\mu}_{i} = \dfrac{1}{N_{t,i}}\sum\limits_{\mathbf{f} \in \mathbf{F}^{i}}{\mathbf{f}},~i = [1,2],
    \label{eq:mean_fingerprint}
\end{equation}
\begin{equation}
    \mathbf{S}_{w} = \sum\limits_{\mathbf{f} \in \mathbf{F}^{1}}{(\mathbf{f} - \mathbf{\mu}_{1})(\mathbf{f} - \mathbf{\mu}_{1})^{T}} + \sum\limits_{\mathbf{f} \in \mathbf{F}^{2}}{(\mathbf{f} - \mathbf{\mu}_{2})(\mathbf{f} - \mathbf{\mu}_{2})^{T}},
    \label{eq:within_scatter_matrix}
\end{equation}
and $\mathbf{S}_{w}$ is the within-class scatter matrix \cite{dhsPC}.
\subsubsection{Principle Component Analysis\label{sec:PCA}}%
PCA is an alternate feature space transformation that reduces dimensionality for the purpose of finding those components most useful for representing the data \cite{dhsPC}. The PCA transformation is a series of projections that are mutually uncorrelated and ordered, from largest to smallest, in variance; thus, the principal axis is aligned along the direction associated with the largest variance~\cite{Hastie}. The principal components are the eigenvectors of the covariance matrix as calculated by,
\begin{equation}
	\Sigma_{\mathbf{\bar{F}}} = \frac{1}{N_{\tau}}\mathbf{\bar{F}}\cdot \mathbf{\bar{F}}^T,
	\label{eq:covariance}
\end{equation}	
where $\mathbf{\bar{F}}$ is a matrix defined as:
\begin{equation}
\resizebox{.91\columnwidth}{!}{$
    \mathbf{\bar{F}} = \mathbf{F} - \begin{bmatrix}
   \mu_{1}^{c_1},
   \mu_{2}^{c_1},
   \dots,
   \mu_{N_{t,1}}^{c_1},
   \dots,
   \mu_{1}^{c_2},
   \dots,
   \mu_{N_{t,2}}^{c_2}
 \end{bmatrix}_{N_{\tau}\times N_{f}}^{T},$}
	\label{eq:RFF_Matrix}
\end{equation}
where $\mu_j$ is the mean of the $j^{\text{th}}$ RF-DNA fingerprint, and $i$$=$$[1,2]$~\cite{Hastie}. The eigenvectors, i.e., principal components, are,
\begin{equation}
(\Sigma_{\mathbf{\bar{F}}} - \lambda_{e} \mathbf{I})e = 0,
\label{eq:Eigenvectors}
\end{equation}	
where $\mathbf{I}$ is the identity matrix, $e$ is the eigenvectors, and $\lambda_{e}$ the eigenvalues~\cite{watkins}. A solution to \eqref{eq:Eigenvectors} is found by solving the characteristic equation,
	\begin{equation}
		|\Sigma_{\mathbf{\bar{F}}} - \lambda_{e} \mathbf{I} | = 0,
		\label{eq:Characteristic}
	\end{equation}
where $|\bullet|$ denotes the determinant~\cite{watkins}. The solution to \eqref{eq:Characteristic} is the eigenvectors for the covariance matrix $\Sigma_{\mathbf{\bar{F}}}$ and the principal components are obtained by ordering the eigenvectors, with respect to their associated eigenvalues, from largest to smallest. There are a total of $N_{f}$ eigenvectors and $N_{r}$ of them are retained to reduce the dimensionality of the RF-DNA fingerprints. Prior to ID verification, the RF-DNA-fingerprints are projected into the PCA defined space by,
\begin{equation}
	\mathbf{F}_{w} = \mathbf{\bar{F}}\cdot\mathbf{P}_{N_{r}},
	\label{eq:PCAProjection}
\end{equation}
where $\mathbf{P}_{N_{r}}$ is a matrix comprised of the first $N_{r}$ eigenvectors associated with the $N_{r}$ largest eigenvalues.
\subsubsection{Neighborhood Component Analysis\label{sec:NCA}}%
NCA is a nearest neighbor-based feature selection approach that maximizes the `Leave-One-Out' classification accuracy \cite{Yang_JCP_2012}. 
Let the training set of RF-DNA fingerprints and the corresponding class label $c_{i}$ be defined as,
\begin{equation}
    \mathbf{\mathcal{F}}_{N_{\tau} \times (N_{f}+1)} = \left[f_{i,r},c_{i}\right],
\end{equation}
where $i$$=$$[1,2]$, and $r$$=$$1,2,\dots,N_{f}$. In NCA, an RF-DNA fingerprint is randomly selected from $\mathbf{\mathcal{F}}$ and is designated as the reference $\mathbf{f}_{j}$. The probability that a new RF-DNA fingerprint $\mathbf{f}_{i}$ is assigned the same class label as $\mathbf{f}_{j}$ is,
\begin{equation}
    p_{ij} = \dfrac{k\left(d_{w}(\mathbf{f}_{i},\mathbf{f}_{j})\right)}{\sum\limits_{j=1,j\neq i}^{N_{\tau}}{k\left(d_{w}(\mathbf{f}_{i},\mathbf{f}_{j})\right)}},
    \label{eq:nca_probability}
\end{equation}
where
\begin{equation}
    d_{w}\left(\mathbf{f}_{i},\mathbf{f}_{j}\right) = \sum\limits_{r=1}^{N_{f}}{w_{r}^{2}\left|f_{ir} - f_{jr}\right|},
    \label{eq:nca_distance}
\end{equation}
\begin{equation}
    \Upsilon(z) = \exp\left(-\dfrac{z}{\psi}\right),
    \label{eq:nca_kernel}
\end{equation}
$\Upsilon$ is the kernel function, $\psi$$=$$1$ is the kernel width, and $w_{r}$ is the weight assigned to the $r^{\text{th}}$ feature. The goal is to find the value of each weight $w_{r}$ such that the chosen subset of RF-DNA fingerprints produces the highest nearest-neighbor classification accuracy \cite{Yang_JCP_2012}. The weight vector $\mathbf{w}$, that achieves this goal, is determined by,
\begin{equation}
    \mathbf{\hat{w}} = \operatorname*{arg\,min}_{w}\left\{\dfrac{1}{N_{\tau}}\sum\limits_{i=1}^{N_{\tau}}{\sum\limits_{j=1,j\neq i}^{N_{\tau}}{p_{ij}l\left(c_{i},c_{j}\right)}} + \lambda_{R}\sum\limits_{r=1}^{N_{f}}{w_{r}^{2}}\right\},
    \label{eq:nca_weight_vector}
\end{equation}
where,
\begin{equation}
    l\left(c_{i},c_{j}\right) = \begin{cases} 
    1 & \text{if } c_{i} \neq c_{j} \\
    0       & \text{otherwise}
    \end{cases},
\end{equation}
is the loss function, and $\lambda_{R}$ is the regularization parameter, which results in most of the weights in $\mathbf{\hat{w}}$ being set to zero. The RF-DNA fingerprint features associated with the $N_{r}$ largest weights are retained and the remainder discarded.
\subsubsection{Probability Of Error \& Average Correlation Coefficient\label{sec:POEACC}}%
POEACC is the weighted sum of the Probability Of Error (POE) and Average Correlation Coefficient (ACC) feature selection techniques. The $r^{\text{th}}$ RF-DNA fingerprint feature is ranked as,
\begin{equation}
    {w}_{r} = w_{\rho}\bar{\rho} + w_{\alpha}\alpha,
    \label{eq:poeacc_weights}
\end{equation}
where $w_{\rho}$ and $w_{\alpha}$ are weights that sum to 1 \cite{MucciardiITC}. The POE values are normalized according to,
\begin{equation}
	\bar{\rho}_{r} = \frac{\rho_r - \rho_{\text{min}}}{\rho_{\text{max}} - \rho_{\text{min}}}.
	\label{eq:poe_norm}
\end{equation}
where $r=1,\dots,N_{f}$. The ACC is given by,
\begin{equation}
    \alpha_{r,q} = \left| \frac{\Sigma_{r,q}}{\sigma_{r}\sigma_{q}}\right|,
		\label{eq:correlation}
\end{equation}
where $r$ and $q$ are a pair of features, $\Sigma_{r,q}$ is the covariance of the training set of RF-DNA fingerprints, and $\sigma_{r}$ as well as $\sigma_{q}$ are the standard deviations of the of the $r^{\text{th}}$ and $q^{\text{th}}$ selected feature, respectively \cite{GonzalezThesis}. As with POE, the ACC is normalized prior to selection of the $r^{\text{th}}$ feature, which ensures that $w_{1}$ and $w_{2}$ remain a true measure of importance in \eqref{eq:poeacc_weights}.

In POEACC, the first feature selected is that which results in the smallest normalized POE, i.e., smallest classification error. The second  feature chosen  is  that  which  results  in  the smallest correlation  value between itself and the first selected feature. The third selected feature is that which results in the smallest average correlation coefficient  between  itself  and  that  of the first two when compared with the average  correlation  coefficients for all remaining  features. This continues until all $N_{f}$ features have been assigned a $w_{r}$ value. Selection of the $r^{\text{th}}$ feature is based upon its average correlation value  with respect to those previously chosen \cite{MucciardiITC}. This work only uses POEACC feature selection; thus, values of $w_{1}$$=$$0$ and $w_{1}$$=$$1$ are neglected as they represent feature selection based only upon ACC  or  POE, respectively.
\subsubsection{Bhattacharyya  Coefficient\label{sec:bhat_coeff}}%
The BC facilitates feature selection by providing a measure of the amount of overlap that exists between histograms. This measure provides an indication of the relative closeness of the histograms. The BC for a given RF-DNA fingerprint feature $f_{r}$ is given by,
\begin{equation}
    \mathcal{B}_{r} = \sum\limits_{b=1}^{N_{\mathcal{B}}}{\sqrt{P_{c_{1}}(b)P_{c_{2}}(b)}},
    \label{eq:bhat_coeff}
\end{equation}
where $N_{\mathcal{B}}$ is the number of `bins/buckets' comprising the histograms, $P_{c_{1}}(b)$ is the probability of $b$ for histogram $c_{1}$, and $P_{c_{2}}(b)$ is the probability of $b$ for the histogram of $c_{2}$ \cite{dhsPC,Comaniciu_CVPR_2000}. The histogram $P_{c_{1}}(b)$ is constructed from the $r^{\text{th}}$ feature of the authorized radio whose ID is to be verified while $P_{c_{2}}(b)$ is constructed from the $r^{\text{th}}$ feature associated with the RF-DNA fingerprints of the remaining authorized radios. If $\mathcal{B}_{r}$$=$$0$, then there is no overlap between the histograms associated with feature $f_{r}$. If $\mathcal{B}_{r}$$=$$1$, then the histograms overlap completely for feature $f_{r}$. For the work presented here, each RF-DNA fingerprint is comprised of $N_{f}$$=$$204$ features and a BC is calculated for each. The RF-DNA fingerprint features corresponding to the smallest BC values are kept, as they indicate the least amount of overlap between histograms, and the remainder are discarded.
\subsubsection{$t$-Test\label{sec:t_test}}%
In this work the Welch's $t$-test is calculated for each of the $N_{f}$ RF-DNA fingerprint features. In the Welch's $t$-test the null hypothesis is tested when the two populations have equal means, but unequal variances or sample sizes \cite{Welch_Biometrika_1947}. As with the Student's $t$-test, the Welch's $t$-test assumes the two populations are distributed normally. This makes Welch's $t$-test well-suited to this work, because the: (i) number of RF-DNA fingerprints comprising the data set differs between $c_{1}$ and $c_{2}$, i.e., unequal sample sizes, and (ii) channel noise is normally distributed \cite{Ruxton_BE_2006,Derrick_TQMP_2016}. For the Welch's $t$-test the test statistic is,
\begin{equation}
    \mathbf{t} = (\mathbf{\mu}_{1}-\mathbf{\mu}_{2})\left(\dfrac{\mathbf{\sigma}_{1}^{2}}{N_{t,1}} + \dfrac{\mathbf{\sigma}_{2}^{2}}{N_{t,2}}\right)^{-1/2},
\end{equation}
where $\mathbf{\mu}_{i}$ is given by \eqref{eq:mean_fingerprint} and $\mathbf{\sigma}_{i}$ is the standard deviation of the $i^{\text{th}}$ RF-DNA fingerprint training set. For the estimated variance, the degrees of freedom $v$ are approximated using the Welch-Satterthwaite equation given by,
\begin{equation}
    v \approx \left(\dfrac{\sigma_{1}^{2}}{N_{t,1}} + \dfrac{\sigma_{2}^{2}}{N_{t,2}}\right)^{2}\left(\dfrac{\sigma_{1}^{4}}{v_{1}N_{t,1}^{2}} + \dfrac{\sigma_{2}^{4}}{v_{2}N_{t,2}^{2}}\right)^{-1}
    \label{eq:welch_satterwaite}
\end{equation}
where $v_{i}$$=$$N_{t,i} - 1$ for $i$$=$$[1,2]$ \cite{Allwood_APStats_2008}. RF-DNA fingerprint features for which the null hypothesis is \emph{rejected} are retained, while the rest are discarded. The retained features are ordered from the smallest to largest probability of observing a $t$ value equal to or larger than that associated with the current value.
\subsubsection{Relief-F\label{sec:relieff}}%
Relief-F extends the Relief algorithm to account for: missing values, noisy data, and more than two classes \cite{Kononenko}. It is for the first two reasons that Relief-F feature selection is used here. Relief-F uses an iterative approach to determine the quality of each RF-DNA fingerprint feature using within feature dimension distances between the selected RF-DNA fingerprint and its $N_{K}$ nearest in-class (a.k.a., \emph{nearest hit}) and $N_{K}$ out-of-class (a.k.a., \emph{nearest miss}) neighbors \cite{Kira01,Kira02}. The weight assigned to the $r^{\text{th}}$ feature is iteratively updated by,
\begin{equation}
	w_{r} = w_{r}' -  \sum\limits_{k=1}^{N_{K}}{\dfrac{\Delta_{r}\left(\tilde{\mathbf{f}},\mathbf{f}_{H}^{k}\right)}{N_{\tau}N_{K}}}  + \sum_{c_{j}\neq c_{i}}{\sum\limits_{k=1}^{N_{K}}{\dfrac{p_{c_{j}}}{1 - p_{c_{i}}}\dfrac{\Delta_{r}\left(\tilde{\mathbf{f}},\mathbf{f}_{M}^{k}\right)}{N_{\tau}N_{K}}}},
	\label{eq:WeightCalc}
\end{equation}
where $w_{r}'$ is the previous weight value of the $r^{\text{th}}$ feature, $\tilde{\mathbf{f}}$ is the RF-DNA fingerprint randomly selected from the set of $N_{\tau}$ training fingerprints, $\mathbf{f}_{H}^{k}$ is one of the $N_{K}$ nearest hits to $\tilde{\mathbf{f}}$, $\mathbf{f}_{M}^{k}$ is one of the $N_{K}$ nearest misses to $\tilde{\mathbf{f}}$, $p_{c_{i}}$ is the prior probability of the class to which $\tilde{\mathbf{f}}$ belongs, $p_{c_{j}}$ is the prior probability of the class to which $\mathbf{f}_{M}^{k}$ belongs,  and
\begin{equation}
    \Delta_{r}\left(\tilde{\mathbf{f}},\mathbf{f}_{\delta}^{k}\right) = \dfrac{\left| \tilde{\mathbf{f}}(r) - \mathbf{f}_{\delta}^{k}(r) \right|}{\text{max}\left\{\tilde{\mathbf{f}}(r),\mathbf{f}_{\delta}^{k}(r)\right\} - \text{min}\left\{\tilde{\mathbf{f}}(r),\mathbf{f}_{\delta}^{k}(r)\right\}},
    \label{eq:relief_diff}
\end{equation}
where $\delta$$=$$[H, M]$ selects nearest hits and misses \cite{Durgabai_IJARCCE_2014,Stief_MMAR_2018}. The retained features are ordered from the largest assigned weight value $w_{r}$ to the smallest. 
\subsection{Support Vector Machines\label{sec:SVM}}%
This section briefly explains the SVM machine learning algorithm. In SVM, the goal is to determine the separating hyper-plane with the largest margin. The larger the margin, then the greater the generality of the classifier \cite{dhsPC}. The optimal hyper-plane is defined by the support vectors, which are the RF-DNA fingerprints that are the most difficult to classify within the training set. Separation of both classes' RF-DNA fingerprints is facilitated through their non-linear mapping into a much higher dimensional space. When sufficiently high, this mapping always facilitates separation of the two classes by a hyper-plane \cite{dhsPC}. 

For the non-separable case, the SVM is defined as,
\begin{equation}
    \underset{\beta,\beta_{0}}{\text{min}}\dfrac{1}{2}||\beta||^{2} + C\sum\limits_{i=1}^{N_{\tau}}{\xi_{i}},~~\xi_{i} \geq 0,
    \label{eq:svm_02}
\end{equation}
\begin{equation}
    \xi_{i} \geq 1 - y_{i}\left(\mathbf{f}_{i}^{T}\beta + \beta_{0}\right),~\forall{i},
\end{equation}
where $\left(\mathbf{f}_{i}^{T}\beta + \beta_{0}\right)$ defines the hyper-plane, $y_{i}$$\in$$[-1,1]$, $\beta$ is a unit vector such that half the margin is $(1/||\beta||)$, $\xi_{i}$ is a `slack' variable that accounts for RF-DNA fingerprints that fall on the wrong side of the margin, and $C$ is a ``cost'' parameter used to tune the function and is set to one for all results presented in Section~\ref{sec:results}. The primal Lagrange function is given as,
\begin{equation}
\begin{aligned}
    L_{P} = \dfrac{1}{2}||\beta||^{2} & +
C\sum\limits_{i=1}^{N_{\tau}}{\xi_{i}} - \sum\limits_{i=1}^{N_{\tau}}{\mu_{i}\xi_{i}} \\
& - C\sum\limits_{i=1}^{N_{\tau}}{\alpha_{i}\left[y_{i}(\mathbf{f}_{i}^{T}\beta + \beta_{0}) - (1 - \xi_{i})\right]},
\end{aligned}
\label{eq:primal}
\end{equation}
which is minimized with respect to $\beta$, $\beta_{0}$, and $\alpha_{i}$. The derivatives of \eqref{eq:primal} are set to zero to obtain,
\begin{align}
    \beta &= \sum\limits_{i=1}^{N_{\tau}}{\alpha_{i}y_{i}\mathbf{f}_{i}}, \label{eq:dev_01}
\end{align}
where $\sum_{i=1}^{N_{\tau}}{\alpha_{i}y_{i}}$$=$$0$, and $\alpha_{i}$$=$$ C - \mu_{i}$. Substituting \eqref{eq:dev_01} into \eqref{eq:primal} results in the Lagrangian dual objective function,
\begin{equation}
    L_{D} = \sum\limits_{i=1}^{N_{\tau}}{\alpha_{i}} - \dfrac{1}{2}\sum\limits_{i=1}^{N_{\tau}}{\sum\limits_{j=1}^{N_{\tau}}{\alpha_{i}\alpha_{j}y_{i}y_{j}\Upsilon\left({\mathbf{f}},{\mathbf{f}}^{\prime}\right)}},
    \label{eq:dual}
\end{equation}
where $\Upsilon\left({\mathbf{f}},{\mathbf{f}}^{\prime}\right)$ maps the RF-DNA fingerprints into the higher dimensional space. In this work, the Radial Basis Function (RBF) is used for this non-linear mapping, which is given by,
\begin{equation}
    \Upsilon\left({\mathbf{f}},{\mathbf{f}}^{\prime}\right) = \exp\left(-\zeta ||{\mathbf{f}} - {\mathbf{f}}^{\prime}||^{2} \right),
    \label{eq:radial_basis_function}
\end{equation}
where $\zeta$ is a positive constant, ${\mathbf{f}}$ and ${\mathbf{f}}^{\prime}$ are two RF-DNA fingerprints \cite{Hastie}. The solutions to \eqref{eq:primal} and \eqref{eq:dual} are found by minimizing $\beta$, $\beta_{0}$, and $\alpha_{i}$ and are given by,
\begin{equation}
    \hat{\beta} = \sum\limits_{i=1}^{N_{\tau}}{\hat{\alpha}_{i}y_{i}\mathbf{f}_{i}},
\end{equation}
where the support vectors $\hat{\beta}$ are the RF-DNA fingerprints for which $\hat{\alpha}_{i}$$\neq$$0$. Support vectors on the edge of the margin have values of $0$$\leq$$\hat{\alpha}_{i}$$\leq$$C$ and $\xi_{i}$$=$$0$. All remaining support vectors have $\hat{\alpha}_{i}$$=$$C$ and $\xi_{i}$$>$$0$. Based upon these constraints and the Karush-Kuhn-Tucker conditions in \cite{Hastie}, then the SVM decision is,
\begin{equation}
    \hat{S}\left(\mathbf{f}\right) = \text{sign}\left[\mathbf{f}\hat{\beta} + \hat{\beta}_{0}\right],
\end{equation}
where `$\text{sign}[\bullet]$' assigns a $-1$ or $1$ to an RF-DNA fingerprint based upon its location with respect to the margin.

As in \cite{Merchant_JSTSP_2018,KandahiThings2019}, radio ID verification is implemented using a two class classifier. One class represents the authorized radio whose  ID  is  to  be  verified, while the second serves to represent all other radios including rogues. During SVM training, this second class is represented using the RF-DNA fingerprints of the remaining authorized radios. 
\subsection{SVM Model Selection}\label{sec:select_svm}%
This section provides an explanation of the process developed to select the SVM model `best' suited to maximize ID verification and rogue radio rejection performance. One of the challenges in ID verification is the selection of a model that is well-suited to not only verification of the authorized radio's ID, but also rejection of rogue radios masquerading as the authorized radio without having access to their RF-DNA fingerprints during model development. The selection of the ``best'' SVM model, i.e., the one that achieves the highest simultaneous ID verification and rogue rejection performance, is predicated on achieving a TVR$\geq$90\% for the authorized radio, whose ID is to be verified, and an FVR$\leq$10\% for the remaining known radios that serve as representative rogues. Thus, if either of these two benchmarks is not satisfied for the selected SVM model, then that SVM model is removed from consideration. If there is no SVM model that satisfies this TVR requirement for a given authorized radio at a particular SNR across all $N_{r}$ retained feature sets, then the SVM model that achieves the highest TVR value for that authorized radio is selected.
\begin{figure}[!t]
    \centering
	\subfigure[Margin PMFs for a model selected for ID verification.]{\label{fig:pmf_chosen}
	\includegraphics[width=\columnwidth]{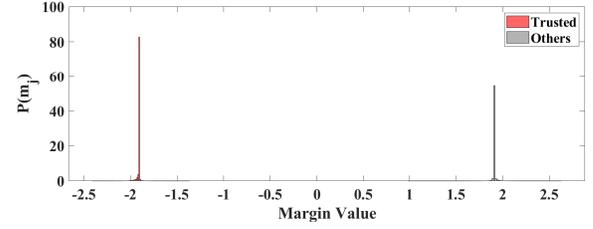}}
	\subfigure[Margin PMFs for a model \emph{not} selected for ID verification.] {\label{fig:pmf_reject}
	\includegraphics[width=\columnwidth]{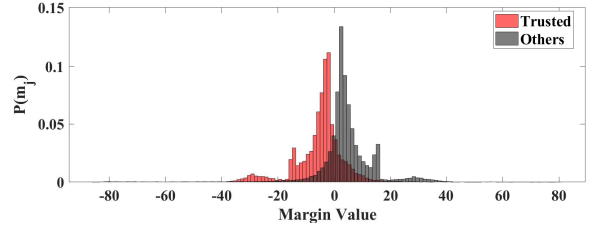}}
	\caption[]{Representative PMFs of the margin values using two different SVM models generated for the same authorized radio and SNR, i.e., a different set of retained RF-DNA fingerprint features are used.}
	\vspace{-5mm}
	\label{fig:pmf_examples}	
\end{figure}
For SVM models that satisfy the TVR$\geq$90\% requirement, the margin is calculated for all of the RF-DNA fingerprints of the authorized radio being verified and the remaining authorized radios, designated herein in as `others', across all noise realizations at the selected SNR and $N_{r}$ features. The margin is calculated by,
\begin{equation}
    m = 2yf\left(\mathbf{f}_{1 \times N_{r}}\right),
    \label{eq:margin}
\end{equation}
\begin{equation}
    f\left(\mathbf{f}_{1 \times N_{r}}\right) = \sum\limits_{j=1}^{N_{\tau}}{\hat{\alpha}_{j}y_{j}G\left(\hat{\beta}_{j},\mathbf{f}_{1 \times N_{r}}\right)} + \hat{\beta}_{0},
    \label{eq:svm_score}
\end{equation}
where $\left(\hat{\alpha}_{1}, \dots, \hat{\alpha}_{N_{T}},\hat{\beta}_{0} \right)$ are the estimated SVM parameters, $y_{j}$$\in$$[-1,1]$, and $G\left(\hat{\beta}_{j},\mathbf{f}_{1 \times N_{r}}\right)$ is the
dot product between the selected RF-DNA fingerprint and the support vectors \cite{Christianini}. Following calculation of the all of the margin values, the Probability Mass Function (PMF) is generated for the positive, i.e., the authorized radio, and negative, i.e., the others, classes. The mean and variance of each PMF is calculated along with the BC value using \eqref{eq:bhat_coeff}. At a given SNR, the SVM model resulting in the largest distance between the PMFs' mean values as well as the smallest BC and variance values is selected as the ``best'' model for ID verification and rogue rejection using the prescribed set of $N_{r}$ features. Figure~\ref{fig:pmf_examples} provides a representative illustration of the two margin PMFs for two different SVM models. In this case, the SVM models are generated for the same authorized radio at the same SNR, but using RF-DNA fingerprints comprised of differing sets of retained features in both those selected and number. The SVM model that resulted in the PMFs shown in Fig.~\ref{fig:pmf_chosen} would be selected for ID verification and rogue radio rejection, while the SVM model associated with Fig.~\ref{fig:pmf_reject} would not.
\subsection{ID Verification Approach}{\label{sec:id_approach}}%
As stated in Sect.~\ref{sec:SOI}, a total of $N_{D}$$=$$18$ WiMAX MS radios are used in this work; thus, ID verification with rogue radio rejection assessment is facilitated by dividing the eighteen radios into two sets: (i) an authorized set of six radios and (ii) a set of twelve to serve as rogue radios. Selection of the six authorized radios was done randomly, while ensuring that each radio is designated as an authorized radio once and a rogue twice. The result is three separate trials of six authorized and twelve rogue radios. The composition of each trial's authorized radios is given in Table~\ref{tbl:WiMAX_Trials} and is consistent with that of \cite{Reising_InfoSec_2015} to enable comparative assessment. For example, to verify the ID of MS63A7, in Trial \#1, then class one of the SVM is trained using MS63A7's RF-DNA fingerprints while class two training is done using the RF-DNA fingerprints of MS63A9, MS66E7, MS6373, MS6387, and MSD905. The remaining twelve radios, i.e., Trial \#2 and \#3 authorized radios, serve as rogues attempting to gain network access by spoofing the digital ID of MS63A7. This process is repeated for each authorized radio of a given trial and across all three trials.
\begin{table}[!h]
\renewcommand{\arraystretch}{1.4}
	\caption{WiMAX Authorized Device Trials \cite{Reising_InfoSec_2015}.}
	 \label{tbl:WiMAX_Trials}
	\centering
	\begin{tabular}{cccc}
	\hline
	{} & \multicolumn{3}{c}{Trial~\#}\\
	\cline{2-4}
	{} & 1 & 2 & 3 \\
	\hline
	\multirow{6}{*}{Digital ID \#}  & MS63A7 & MS637D & MSC2FF \\
	{} & MS63A9 & MS9993 & MSDAC5 \\
	{} & MS66E7 & MSDAB9 & MSDDC7 \\
	{} & MS6373 & MSDAC9 & MSDF5B \\
	{} & MS6387 & MSDADB & MSDF7D \\
	{} & MSD905 & MSDDBF & MSDF65 \\
	\hline
	\end{tabular}
	\vspace{-4mm}
\end{table}
\section{Evaluation}\label{sec:results}%
In this section, the developed ID verification and rogues radio rejection process, which uses feature reduced RF-DNA fingerprints and a two class SVM, is evaluated for its effectiveness in mitigating the threat as described in Sect.~\ref{sec:threat_model}. All ID verification and rogue radio rejection results are generated using SVM models that are developed following the approach described in Sect.~\ref{sec:id_approach}. For a given SNR value, an authorized radio's SVM model is developed using Monte Carlo simulation that is enabled through the use of $N_{z}$$=$$10$ independent, like-filtered AWGN realizations being added to every radio's near-transient responses prior to RF-DNA fingerprint generation. The SVM classifier is trained using $k$$=$$5$-fold cross validation and $N_{b}$$=$$900$ $N_{r}$-dimensional RF-DNA fingerprints for each of the authorized radios; thus, class one and two are represented using 900 and 5,400 RF-DNA fingerprints, respectively. Each feature selection approach, with the exception of LDA, designates the $N_{r}$$\in$$[1, 200]$ top ranked RF-DNA fingerprint features as described in Sect.~\ref{sec:DRA} through Sect.~\ref{sec:relieff}. For each authorized radio, a total of $k$$\times$$N_{z}$$=$$50$ SVM models are generated at each value of $N_{r}$; thus, the model that results in the smallest classification error, across all Monte Carlo trials and $k$-fold steps, is used to represent the selected authorized radio at the particular value of $N_{r}$. The authorized radio's ``best'' SVM model, across all values of $N_{r}$, is chosen using the procedure presented in Sect.~\ref{sec:select_svm}.
\begin{figure*}[!t]
\begin{centering}
\begin{subfigure}[Claimed ID: MS63A7]{\label{fig:all_feat_a}
  \includegraphics[width=0.3\linewidth]{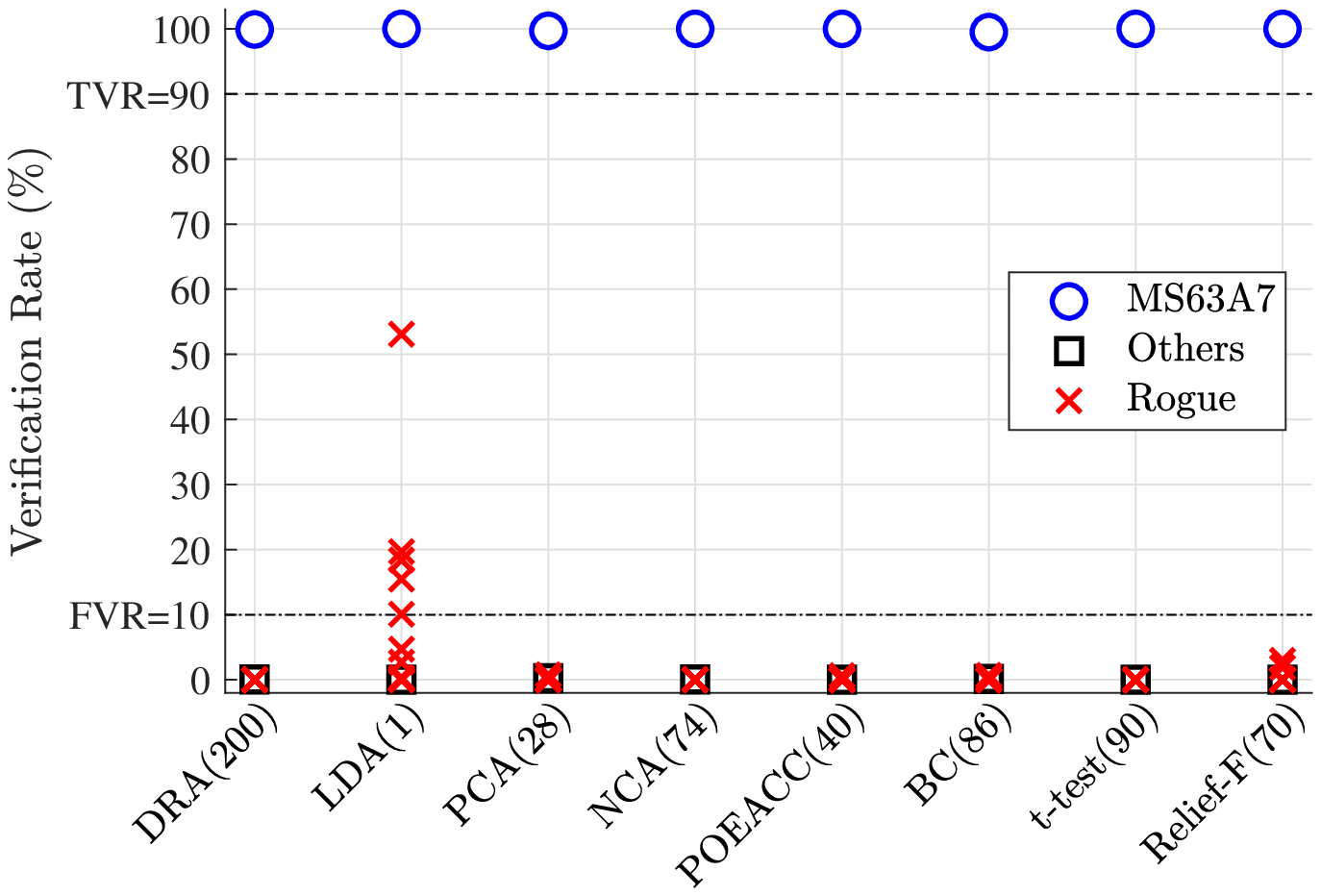}}
\end{subfigure}\hfil 
\begin{subfigure}[Claimed ID: MS63A9]{\label{fig:all_feat_b}
  \includegraphics[width=0.3\linewidth]{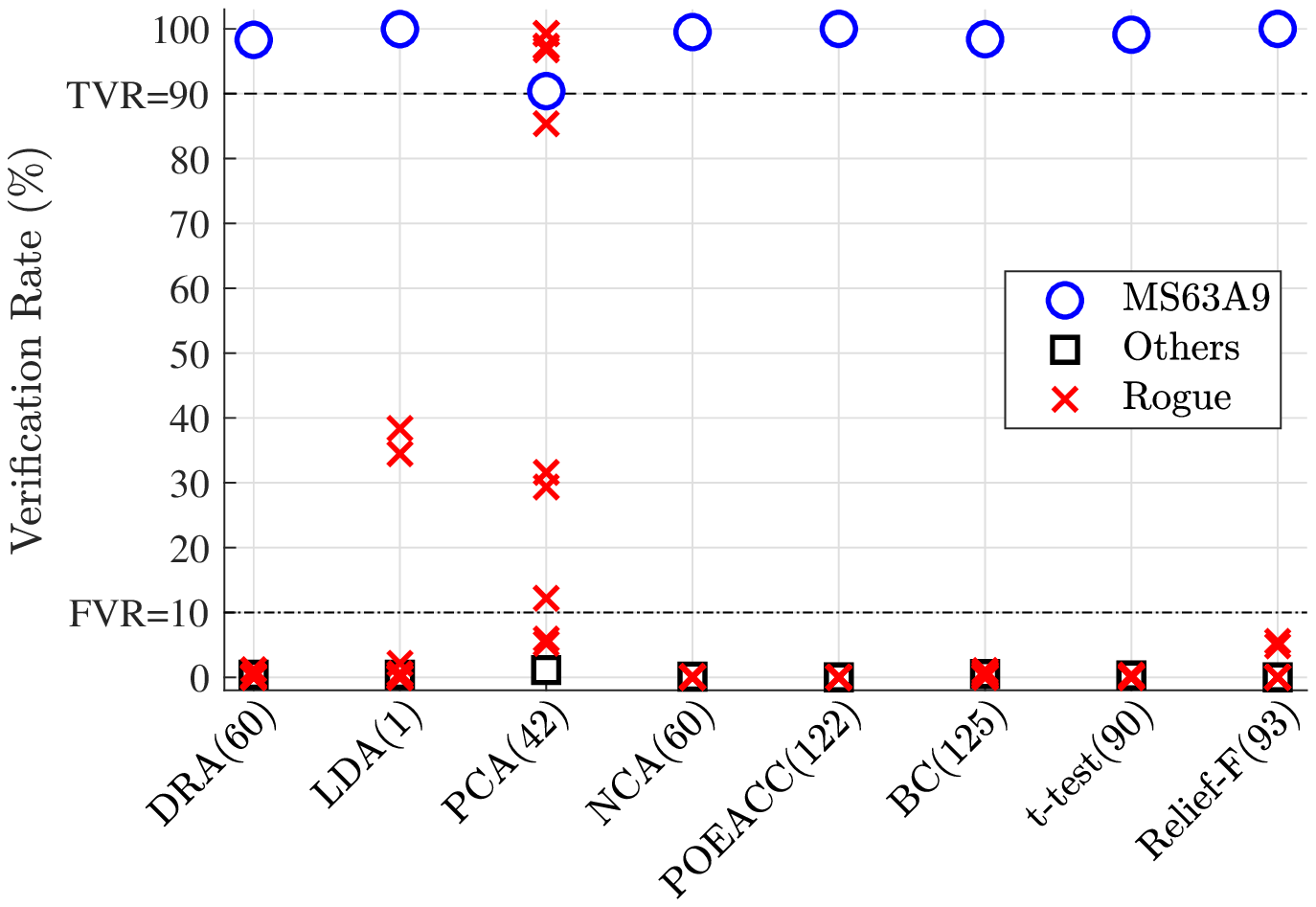}}
\end{subfigure}\hfil 
\begin{subfigure}[Claimed ID: MS66E7]{\label{fig:all_feat_c}
  \includegraphics[width=0.3\linewidth]{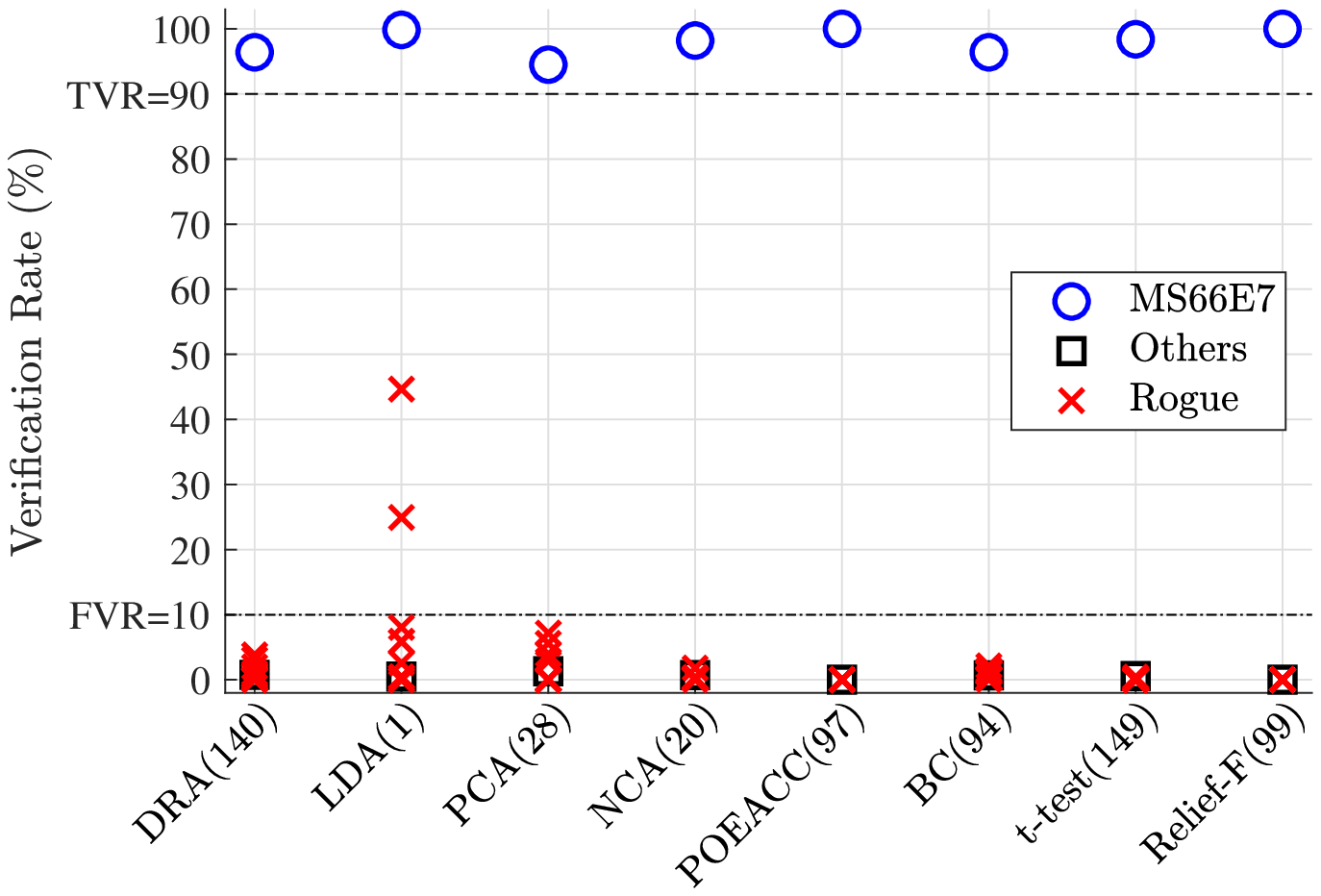}}
\end{subfigure}
\medskip
\begin{subfigure}[Claimed ID: MS6373]{\label{fig:all_feat_d}
  \includegraphics[width=0.3\linewidth]{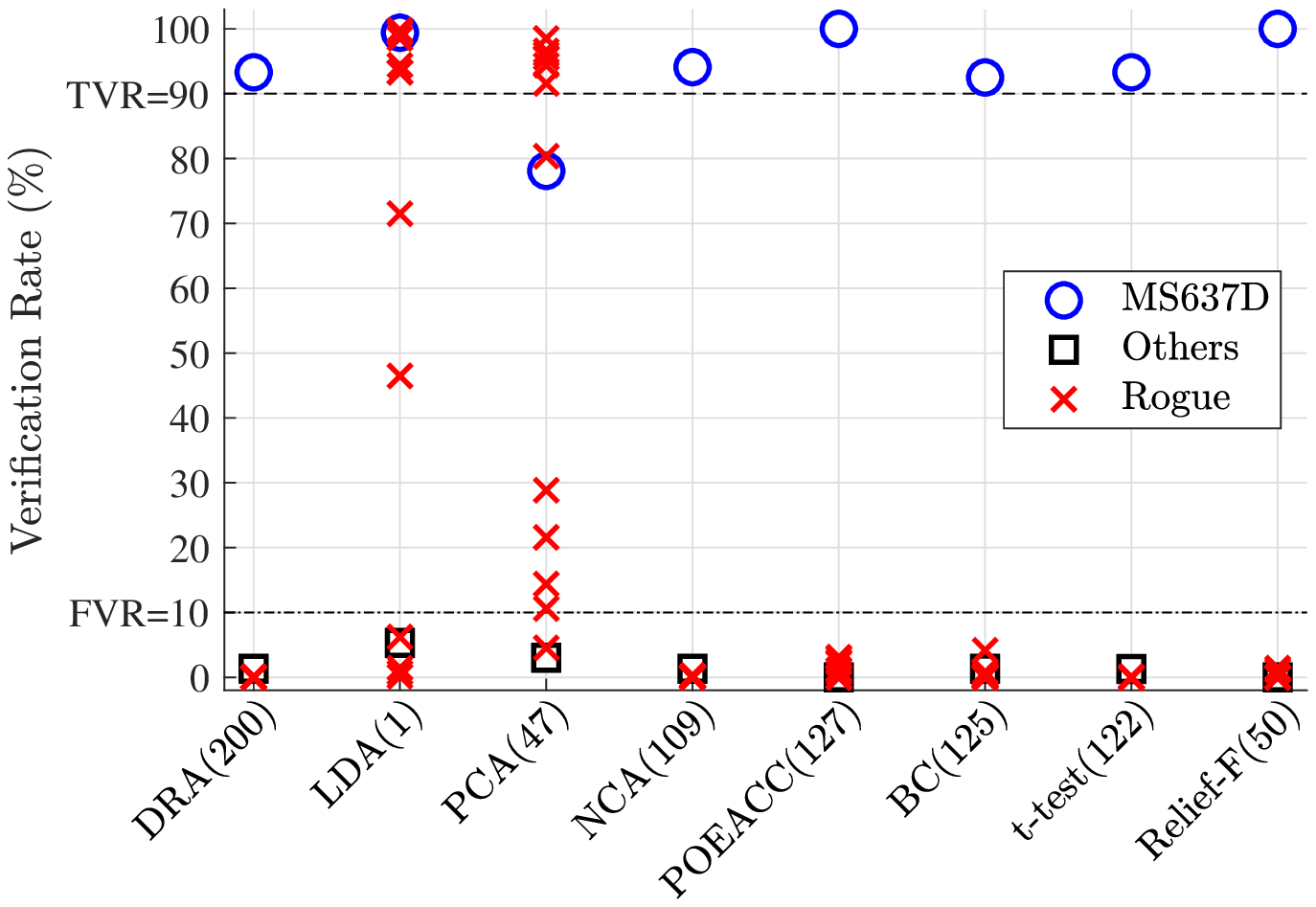}}
\end{subfigure}\hfil 
\begin{subfigure}[Claimed ID: MS6387]{\label{fig:all_feat_e}
  \includegraphics[width=0.3\linewidth]{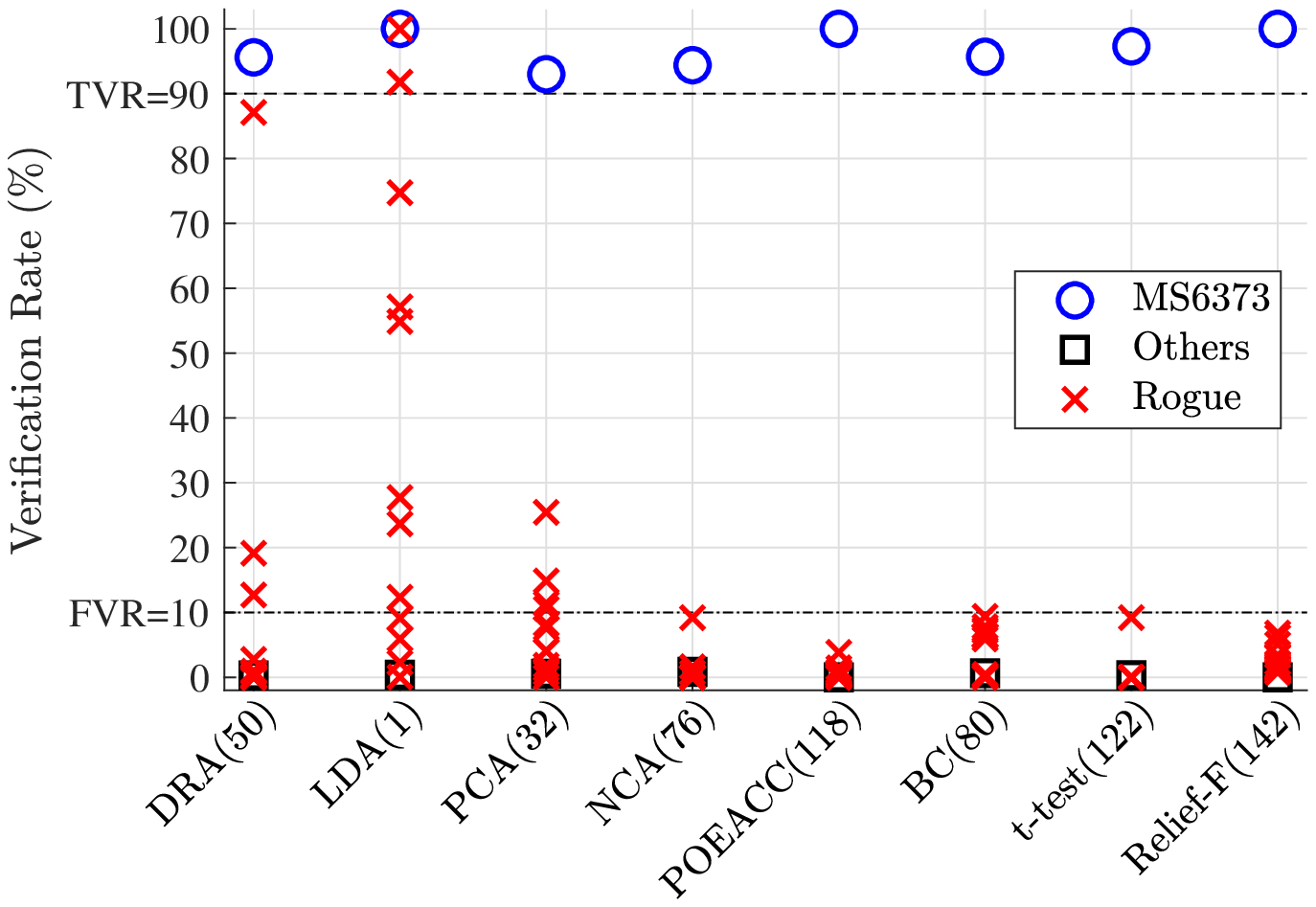}}
\end{subfigure}\hfil 
\begin{subfigure}[Claimed ID: MSD905]{\label{fig:all_feat_f}
  \includegraphics[width=0.3\linewidth]{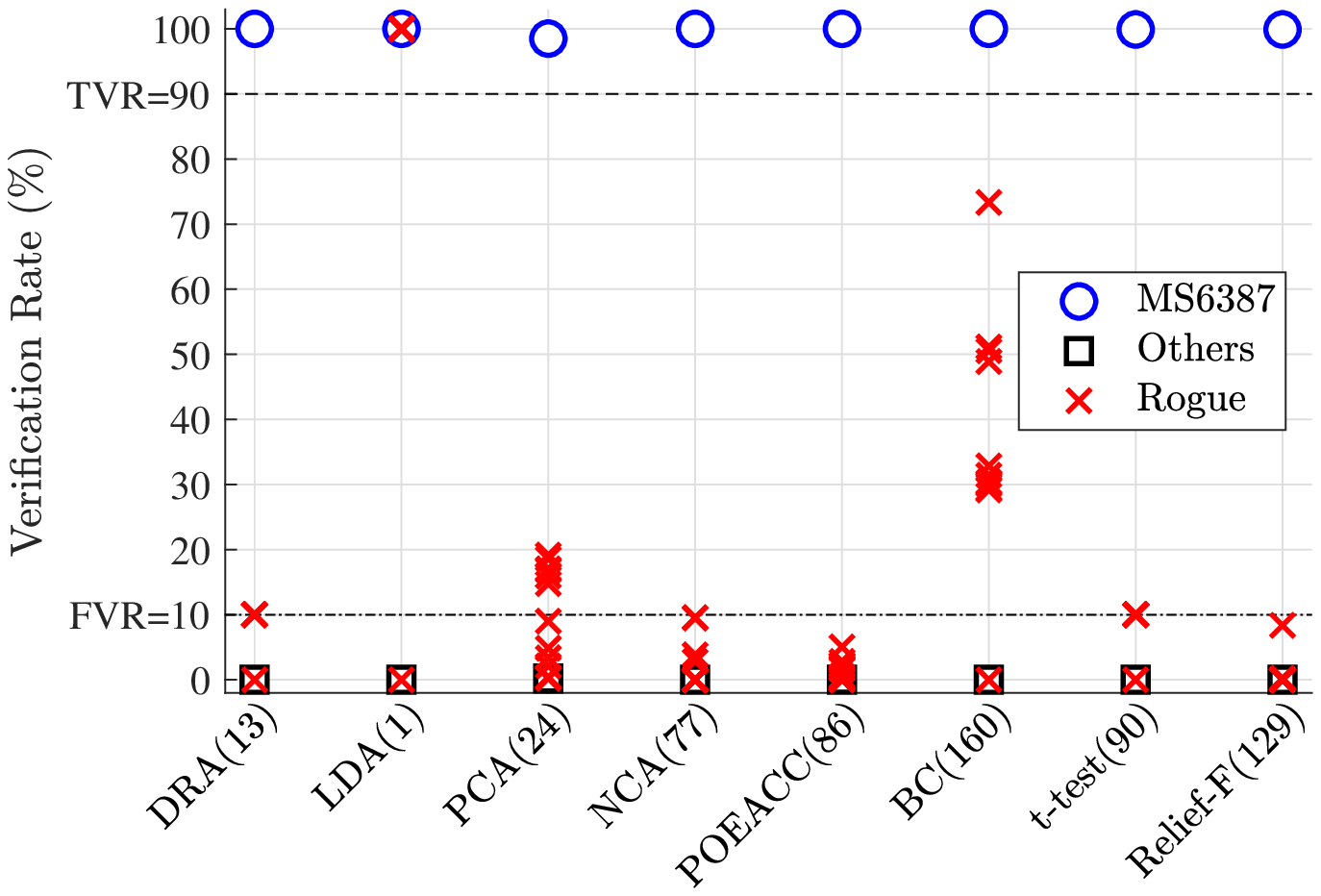}}
\end{subfigure}
\caption{ID verification $({\color{blue}{\circ}})$ and rogue rejection $({\color{red}{\times}})$ performance for the six authorized radios of Trial \#1 using all eight feature selection methods at \underline{SNR$=$21~{dB}}. The number of retained features is in parentheses along the x-axis and ``Others $(\square)$'' indicates the five authorized radios whose ID is not being verified.}
\vspace{-5mm}
\label{fig:all_feat_methods}
\end{centering}
\end{figure*}
\subsection{Evaluation of the Feature Selection Approaches}\label{sec:results_feature_select}%
The purpose of this section is to determine the feature selection method(s) that are most effective in choosing the set and number of retained RF-DNA fingerprint features based on ID verification and rogue rejection performance. For the authorized radio, whose ID is to be verified, the ``best'' ID verification performance is defined as achieving a TVR$\geq$90\%, while the ``best'' rogue rejection performance is achieving an FVR$\leq$10\% for all rogue radios and the remaining authorized radios, i.e., those whose ID is \emph{not} being verified by the selected SVM model. The feature selection method(s) that simultaneously achieves both of these conditions, for all authorized radios, is used for SNR-based analysis. \\
\indent The effectiveness of each feature selection method is assessed using the authorized radios of Trial \#1 at SNR$=$21~{dB}. Selection of this trial and SNR is motivated by the published results in \cite{Reising_InfoSec_2015}, which presents ID verification and rogue radio rejection performance using the exact same: set of $N_{f}$-dimensional RF-DNA fingerprints, noise realizations, authorized and rogue radios, as well as SNR value.\\
\indent Our results are presented in Fig.~\ref{fig:all_feat_methods}, the ID verification for each of the six authorized radios comprising Trial \#1 at SNR$=$21~{dB}. For authorized radio MS63A7, shown in Fig.~\ref{fig:all_feat_a}, a TVR$=$100\% is achieved using RF-DNA fingerprint features selected using all eight feature selection approaches. The other five authorized radios, i.e., MS63A9, MS66E7, MS6373, MS6387, and MSD905, are successfully rejected at a FVR$<$10\% using the features selected by all eight methods. The digital ID spoofing attacks by the rogue radios, i.e., those comprising Trials \#2 and \#3, are successfully rejected at an FVR$<$10\% using: DCA, PCA, NCA, POEACC, BC, $t$-test, and Relief-F selected features. LDA selected RF-DNA features fail to satisfy the FVR$\leq$10\% requirement for five of the twelve total digital ID spoofing attacks. The worst case being a rogue radio that achieves a successful attack rate of $\sim$53\%, i.e., granted network access, when spoofing the digital ID of MS63A7.\\ 
\indent ID verification and rogue radio rejection results for each of the eight feature selection approaches is presented in Fig.~\ref{fig:all_feat_b} for the authorized radio MS63A9. The ID of MS63A9 is successfully verified at a TVR$\geq$90\% using the features designated by all eight selection methods. The other five authorized radios are all successfully rejected at an FVR$\leq$2\% regardless of the feature selection method used. For rogue radio rejection, RF-DNA fingerprints comprised of LDA and PCA selected features fail to achieve the required FVR$\leq$10\% benchmark for all twelve rogue radio attacks. For PCA selected features, two rogues are granted network access at rates of over 96\% and as high as 99\%. This means that these two rogue radios are virtually indistinguishable from the authorized radio, MS63A9, when presenting its digital ID and PCA is used to select the RF-DNA fingerprint features.\\ 
\indent The ID verification and rogue rejection results for MS66E7 are presented in Fig.~\ref{fig:all_feat_c}. The ID of MS66E7 is successfully verified using RF-DNA fingerprint features selected by each of the eight methods described in Sect.~\ref{sec:feat_methods}. The feature reduced RF-DNA fingerprints of the other authorized radios are successfully rejected at an FVR$\leq$3\% for all eight feature selection methods. The rejection performance of the other authorized radios is important to prevent anyone of them from being incorrectly verified as the authorized radio whose ID is being verified. Two of the rogue radios are falsely verified, as MS66E7, at rates of 25\% and 45\% when the retained RF-DNA fingerprint features are chosen using LDA. \\
\begin{figure*}[!t]
\begin{centering}
\begin{subfigure}[Trial \#1 using POEACC.]{\label{fig:poeacc_9dB_a}
  \includegraphics[width=0.3\linewidth]{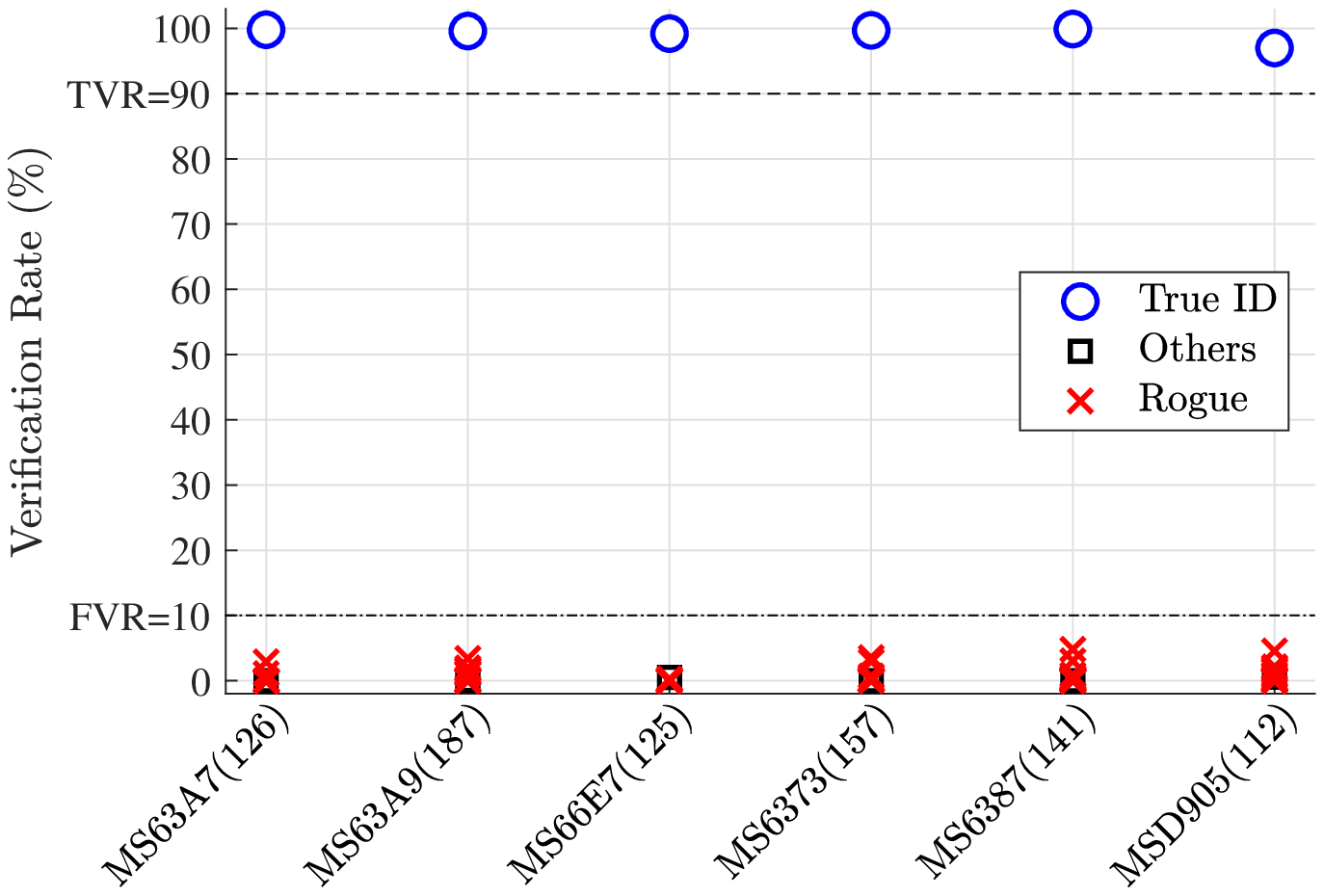}}
\end{subfigure}\hfil 
\begin{subfigure}[Trial \#2 using POEACC.]{\label{fig:poeacc_9dB_b}
  \includegraphics[width=0.3\linewidth]{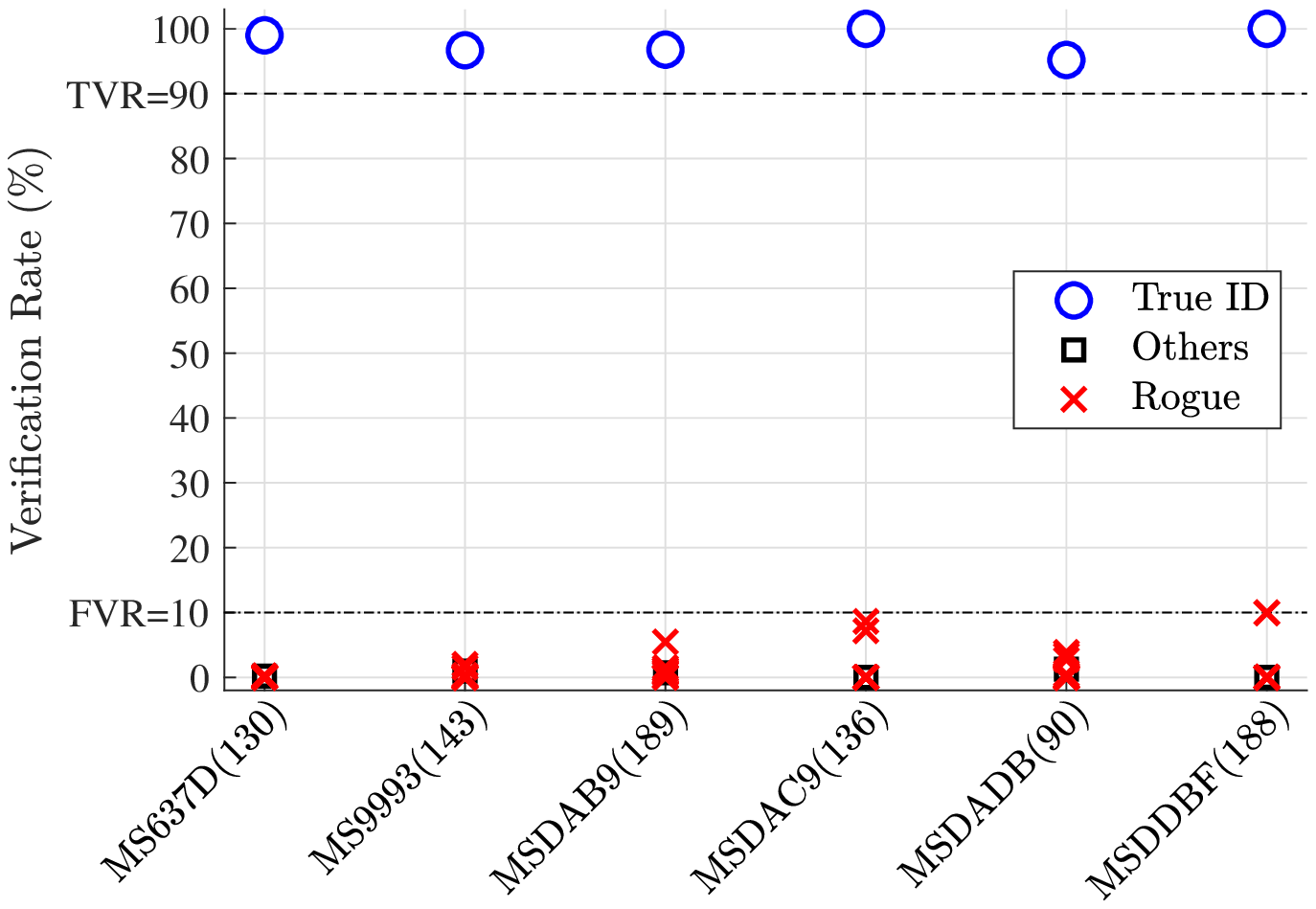}}
\end{subfigure}\hfil 
\begin{subfigure}[Trial \#3 using POEACC.]{\label{fig:poeacc_9dB_c}
  \includegraphics[width=0.3\linewidth]{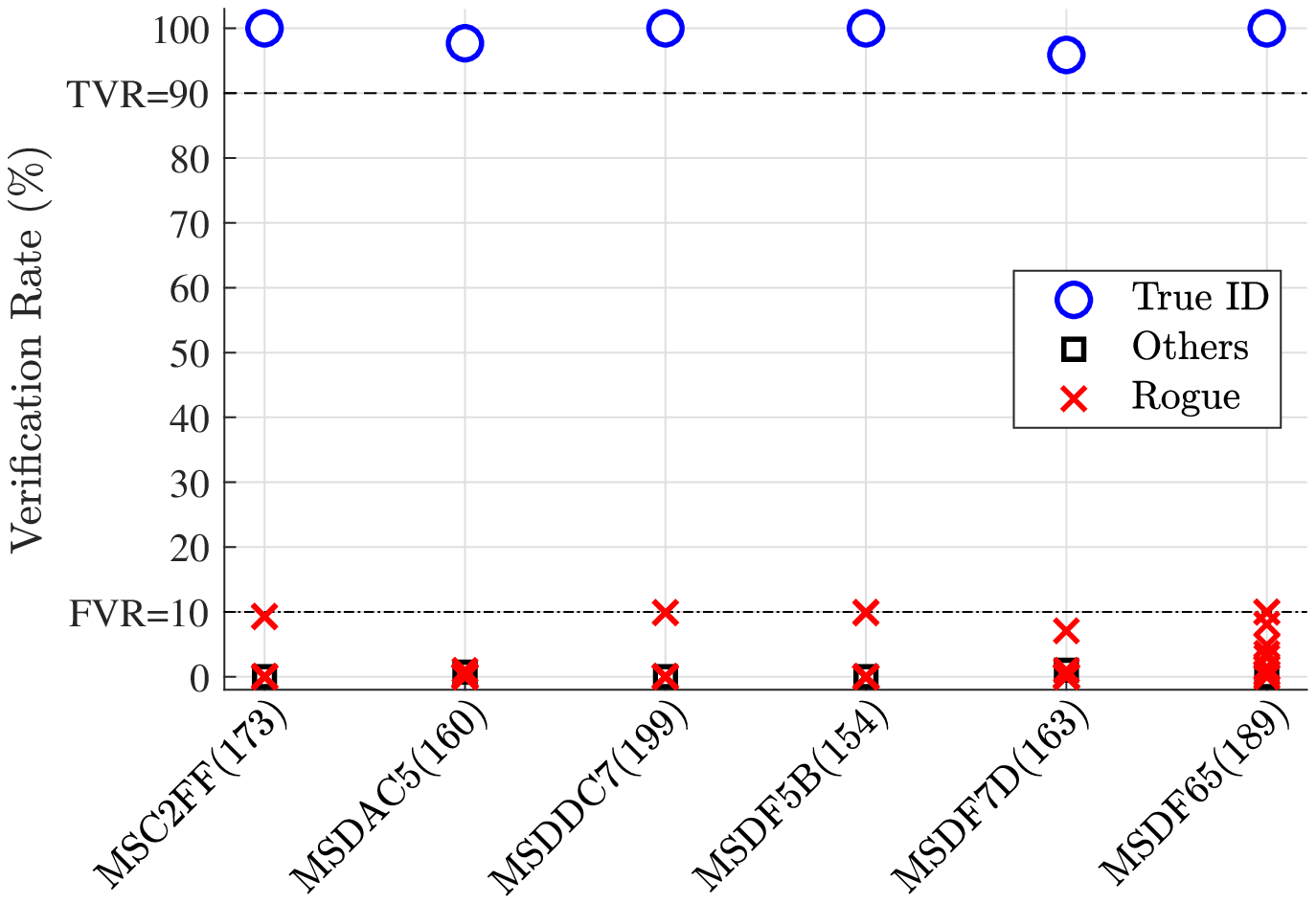}}
\end{subfigure}
\medskip
\begin{subfigure}[Trial \#1 using Relief-F.]{\label{fig:reliefF_9dB_a}
  \includegraphics[width=0.3\linewidth]{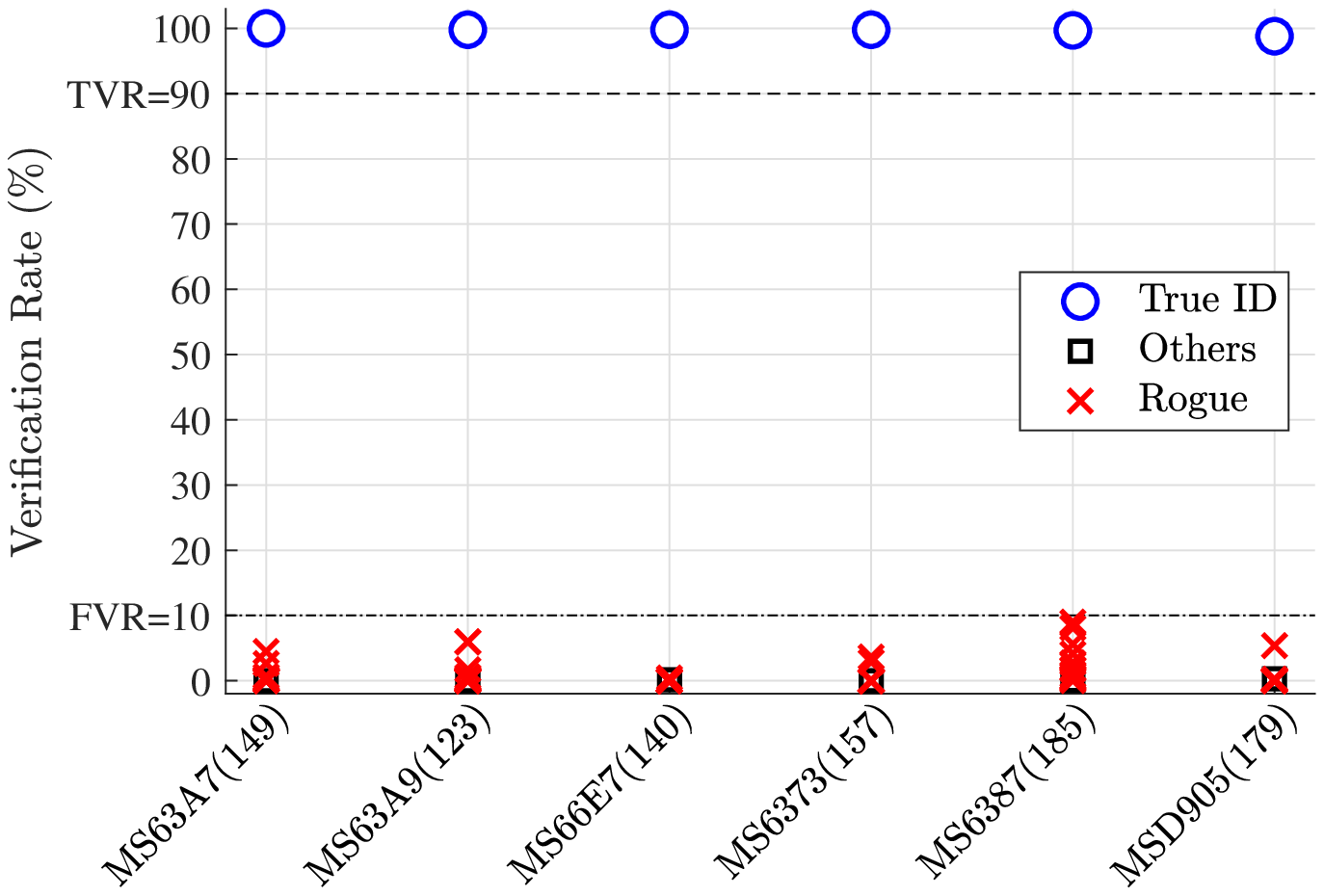}}
\end{subfigure}\hfil 
\begin{subfigure}[Trial \#2 using Relief-F.]{\label{fig:reliefF_9dB_b}
  \includegraphics[width=0.3\linewidth]{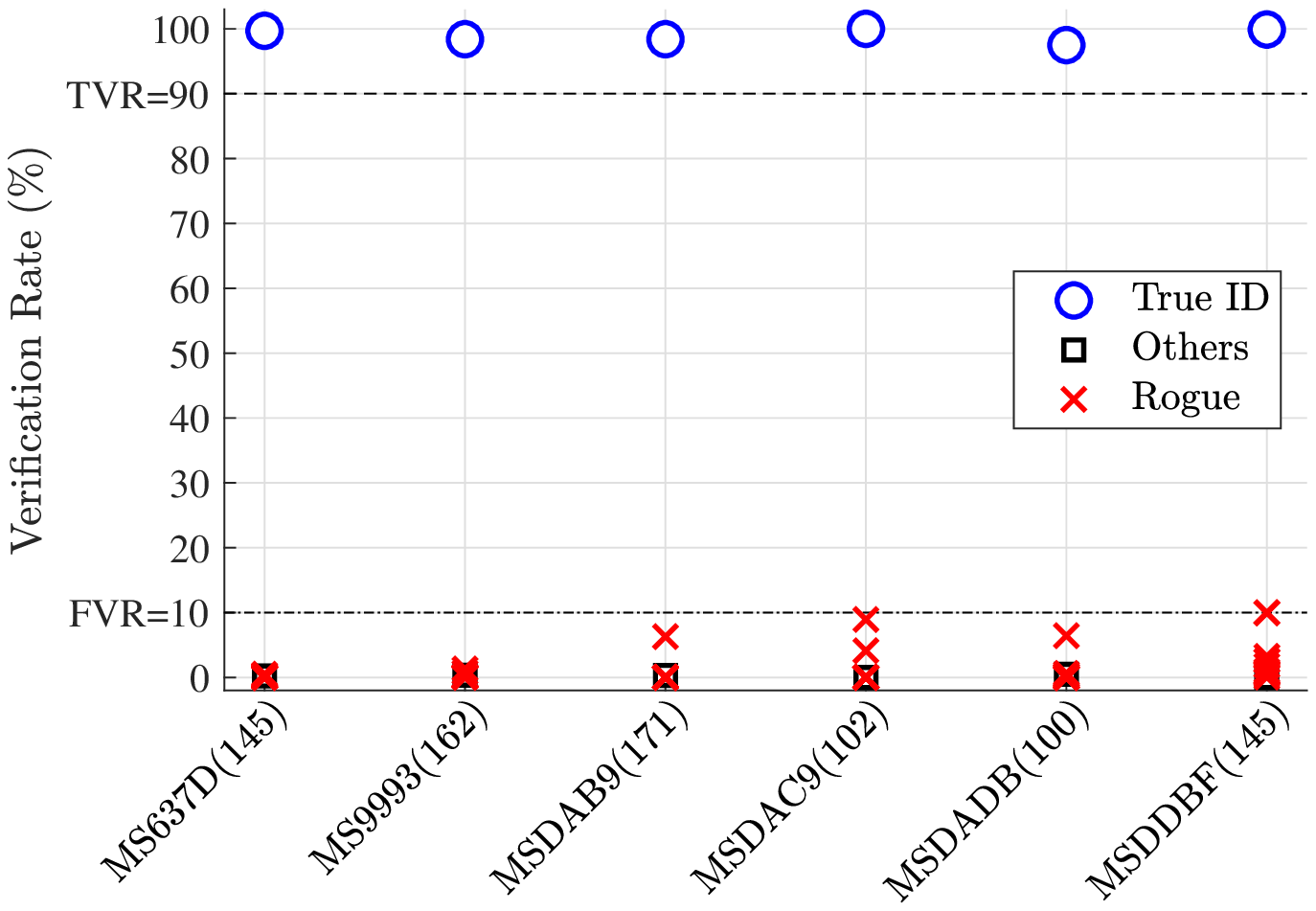}}
\end{subfigure}\hfil 
\begin{subfigure}[Trial \#3 using Relief-F.]{\label{fig:reliefF_9dB_c}
  \includegraphics[width=0.3\linewidth]{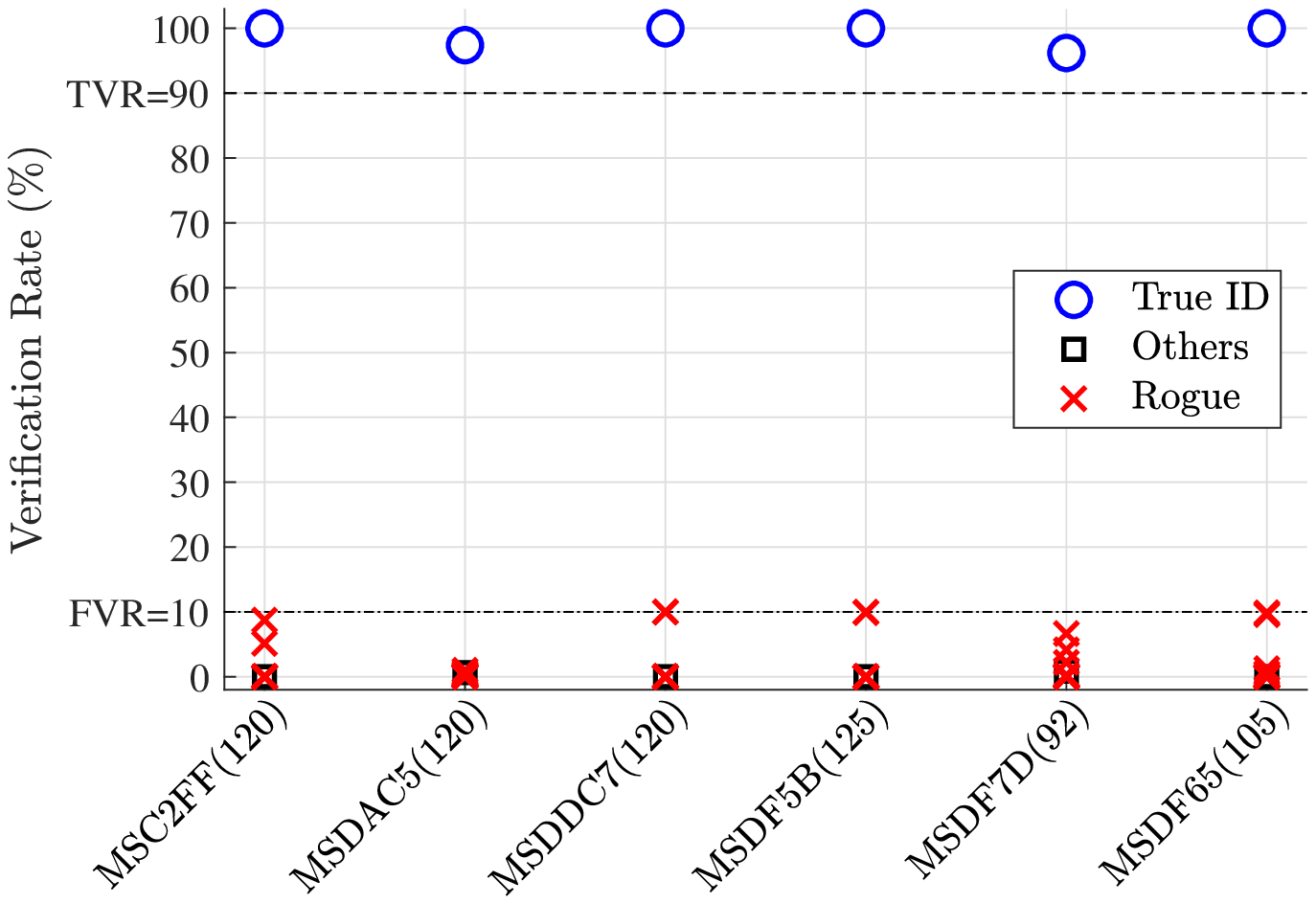}}
\end{subfigure}
\caption{ID verification $({\color{blue}{\circ}})$ and rogue rejection $({\color{red}{\times}})$ performance for the six authorized radios of \emph{all} three trials using the POEACC and Relief-F feature selection methods at \underline{SNR$=$9~{dB}}. The x-axis labels indicate the \emph{claimed} digital ID with the number of retained features in parentheses and ``Others $(\square)$'' indicates the five authorized radios whose ID is not being verified.}
\vspace{-5mm}
\label{fig:poeacc_reliefF_9dB}
\end{centering}
\end{figure*}
\indent The ID verification and rogue rejection performance for authorized radio MS6373 is shown in Fig.~\ref{fig:all_feat_d}. The ID of MS6373 is successfully verified at a TVR$\geq$90\% using RF-DNA fingerprints whose features are selected using the: DRA, LDA, NCA, POEACC, BC, $t$-test, and Relief-F, methods. When using PCA selected RF-DNA fingerprint features, the ID of MS6373 is verified 78\% of the time. The remaining authorized radios are all successfully rejected at an FVR$\leq$7\% when using any one of the eight methods to reduce RF-DNA fingerprint dimensionality. For MS6373, all twelve rogue radio attacks are successfully rejected when using: DRA, NCA, POEACC, BC, $t$-test or Relief-F reduced RF-DNA fingerprints. When using LDA and PCA reduced RF-DNA fingerprints a total of 5 and 6 rogue radios are incorrectly verified as radio MS6373 at a FVR$\geq$90\%, respectively. LDA assumes that the discriminating information is contained in the class means; thus, the poorer rogue rejection performance is attributed to violation of this assumption \cite{dhsPC}. In PCA-based feature selection, it is assumed that the principle components are: (i) linear combinations of the original features comprising the RF-DNA fingerprint(s), (ii) orthogonal, and (iii) associated with the axes of highest variance \cite{dhsPC}. If one or more of these assumptions do not hold, then PCA will fail to select a set of RF-DNA fingerprint features that facilitates separation of one or more rogue radios from the authorized radio whose ID is being verified by the developed SVM model. \\
\indent MS6387 ID verification and rogue rejection performance is presented in Fig.~\ref{fig:all_feat_e}. The ID of MS6387 is successfully verified at a TVR$\geq$90\% using all eight retained RF-DNA fingerprint feature sets. An FVR$\leq$2\% is achieved for the other authorized radios. However, rogue rejection using: DRA, LDA, and PCA selected RF-DNA fingerprint features fails to achieve the required FVR$\leq$10\% benchmark for: 3, 8, and 4 of the twelve rogue attacks, respectively. For LDA, two of the rogue radios' ID's are verified as that of MS6387 at an FVR$=$[92, 100]\%. \\
\indent Figure~\ref{fig:all_feat_f} shows that the ID of MSD905 is verified at a TVR$\geq$98\%, while the other authorized radios are correctly rejected at an FVR$\leq$1\%. For rogue radios spoofing the ID of MSD905, the required FVR$\leq$10\% is achieved using the top ranked RF-DNA fingerprint features selected using five, i.e., DRA, NCA, POEACC, $t$-test, and Relief-F, of the eight methods. The worst case rogue rejection performance is an FVR$=$100\%, which results when using LDA selected RF-DNA fingerprint features. \\
\indent Considering the ID verification results, in Fig.~\ref{fig:all_feat_methods}, across the six authorized radios of Trial \#1, the required TVR$\geq$90\% and FVR$\leq$10\% benchmarks are achieved using $N_{r}$-dimensional RF-DNA fingerprints whose top features are ranked using: NCA, POEACC, $t$-test, and Relief-F. The remaining four feature selection methods, i.e., DRA, LDA, PCA, and BC, fail to satisfy one or both of the requisite benchmarks for at least one of the six authorized radio ID verification scenarios. The worst case occurs when using LDA reduced RF-DNA fingerprints, which fails to achieve the rogue rejection FVR$\leq$10\% requirement for all six authorized radio ID verification scenarios. LDA's poor performance is attributed to the projection itself, because it assumes that the two classes are linearly separable. If the two classes are \emph{not} linearly separable, then LDA cannot discriminate between them \cite{Tharwat_AIComms_2017}. The LDA projection reduces the $N_{f}$$=$$204$-dimensional RF-DNA fingerprints onto one-dimension, which leaves very little, if any, discriminating information for discerning the RF-DNA fingerprints of a rogue from that of the authorized radio.\\
\begin{figure*}[!t]
\begin{centering}
\begin{subfigure}[Trial \#1 at SNR$=$6~{dB}.]{\label{fig:reliefF_6dB_a}
  \includegraphics[width=0.3\linewidth]{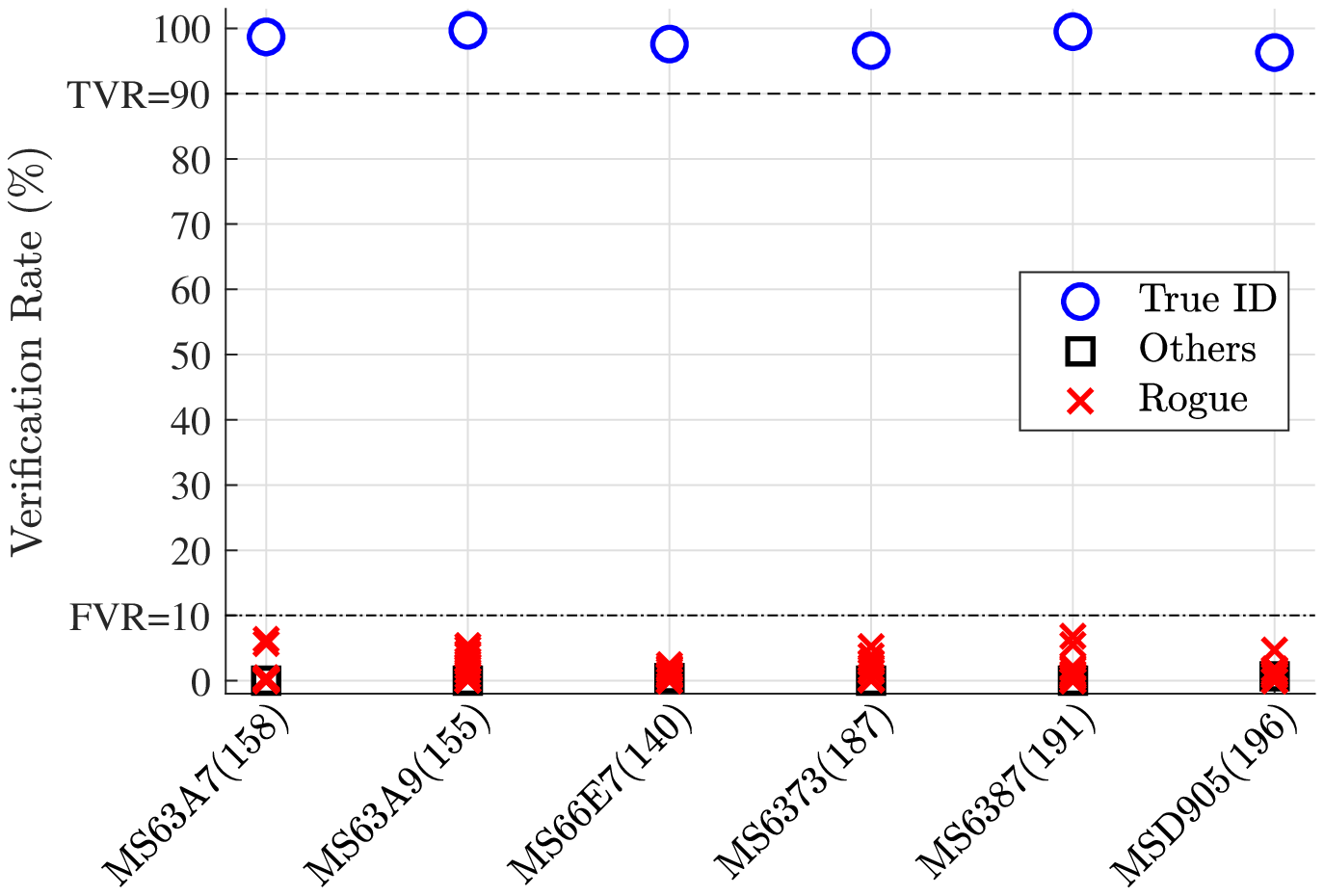}}
\end{subfigure}\hfil 
\begin{subfigure}[Trial \#2 at SNR$=$6~{dB}.]{\label{fig:reliefF_6dB_b}
  \includegraphics[width=0.3\linewidth]{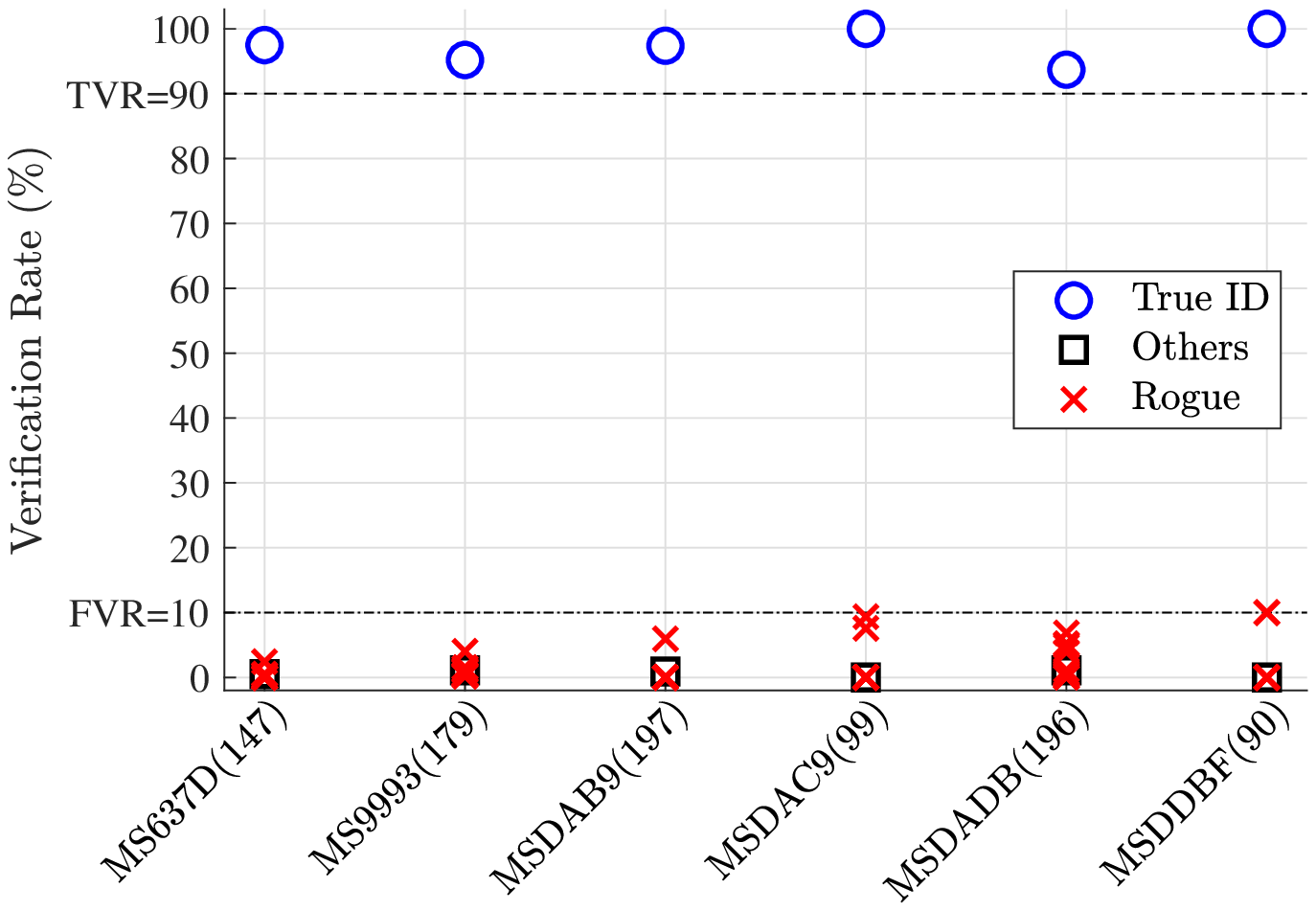}}
\end{subfigure}\hfil 
\begin{subfigure}[Trial \#3 at SNR$=$6~{dB}.]{\label{fig:reliefF_6dB_c}
  \includegraphics[width=0.3\linewidth]{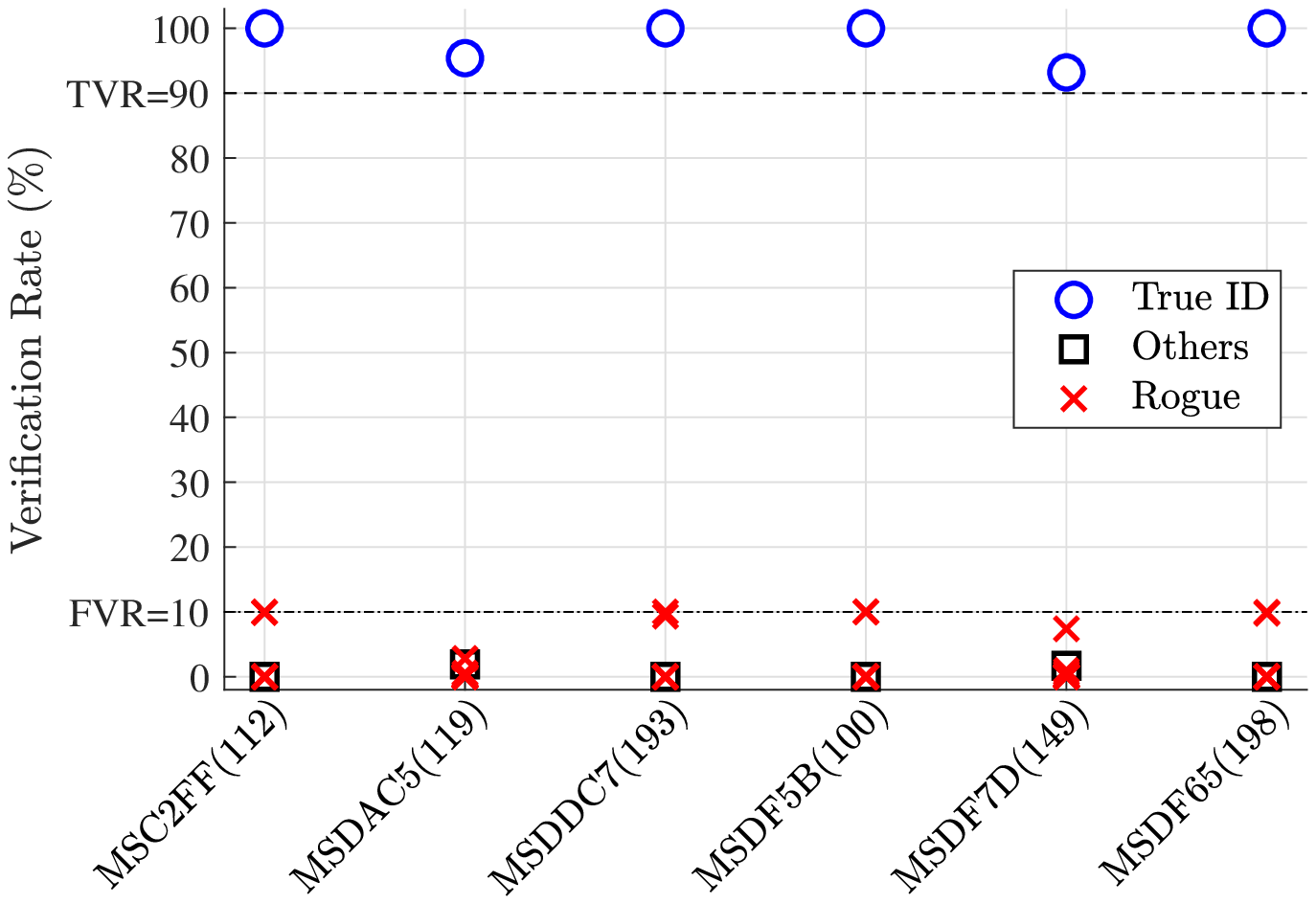}}
\end{subfigure}
\medskip
\begin{subfigure}[Trial \#1 at SNR$=$3~{dB}.]{\label{fig:reliefF_3dB_a}
  \includegraphics[width=0.3\linewidth]{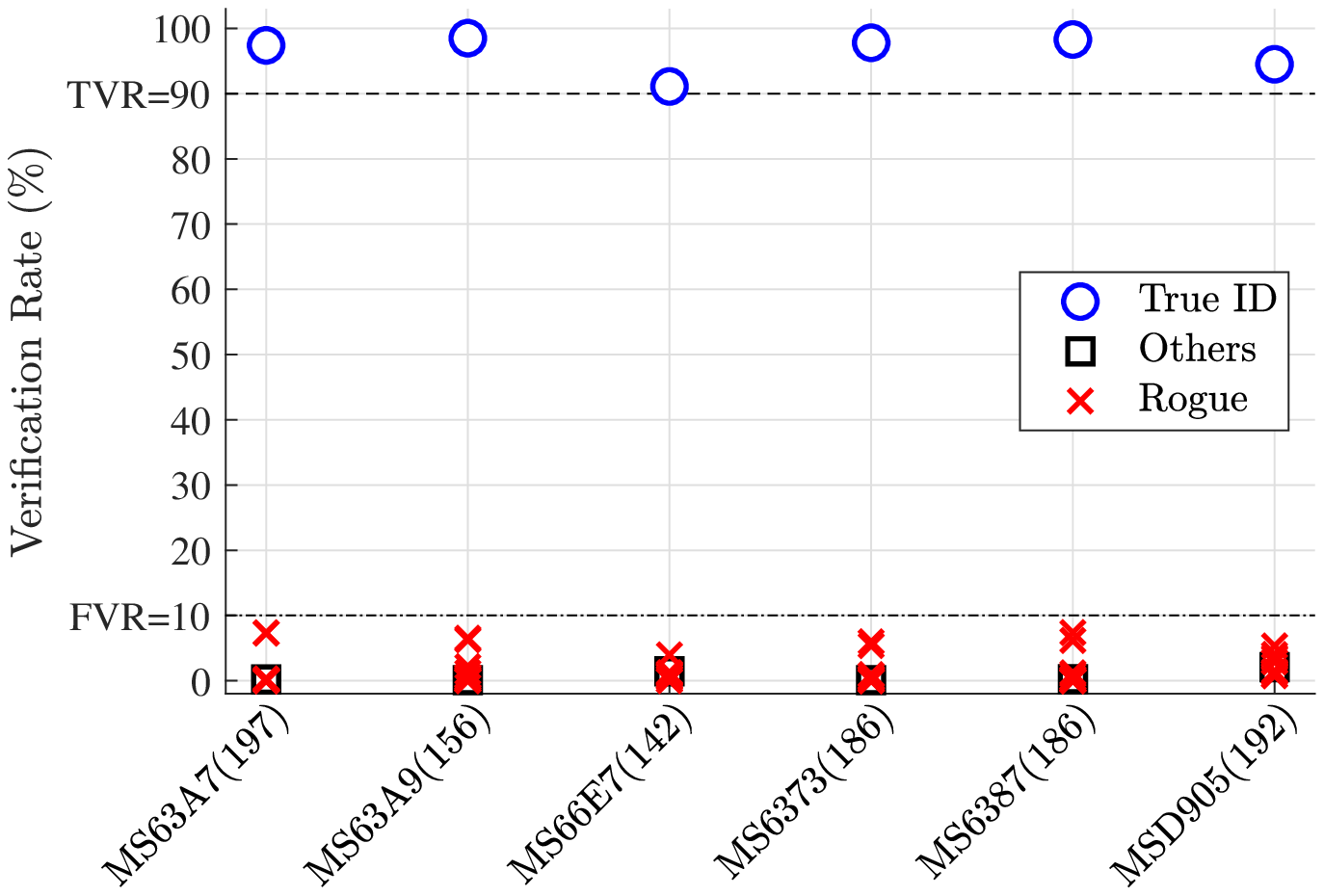}}
\end{subfigure}\hfil 
\begin{subfigure}[Trial \#2 at SNR$=$3~{dB}.]{\label{fig:reliefF_3dB_b}
  \includegraphics[width=0.3\linewidth]{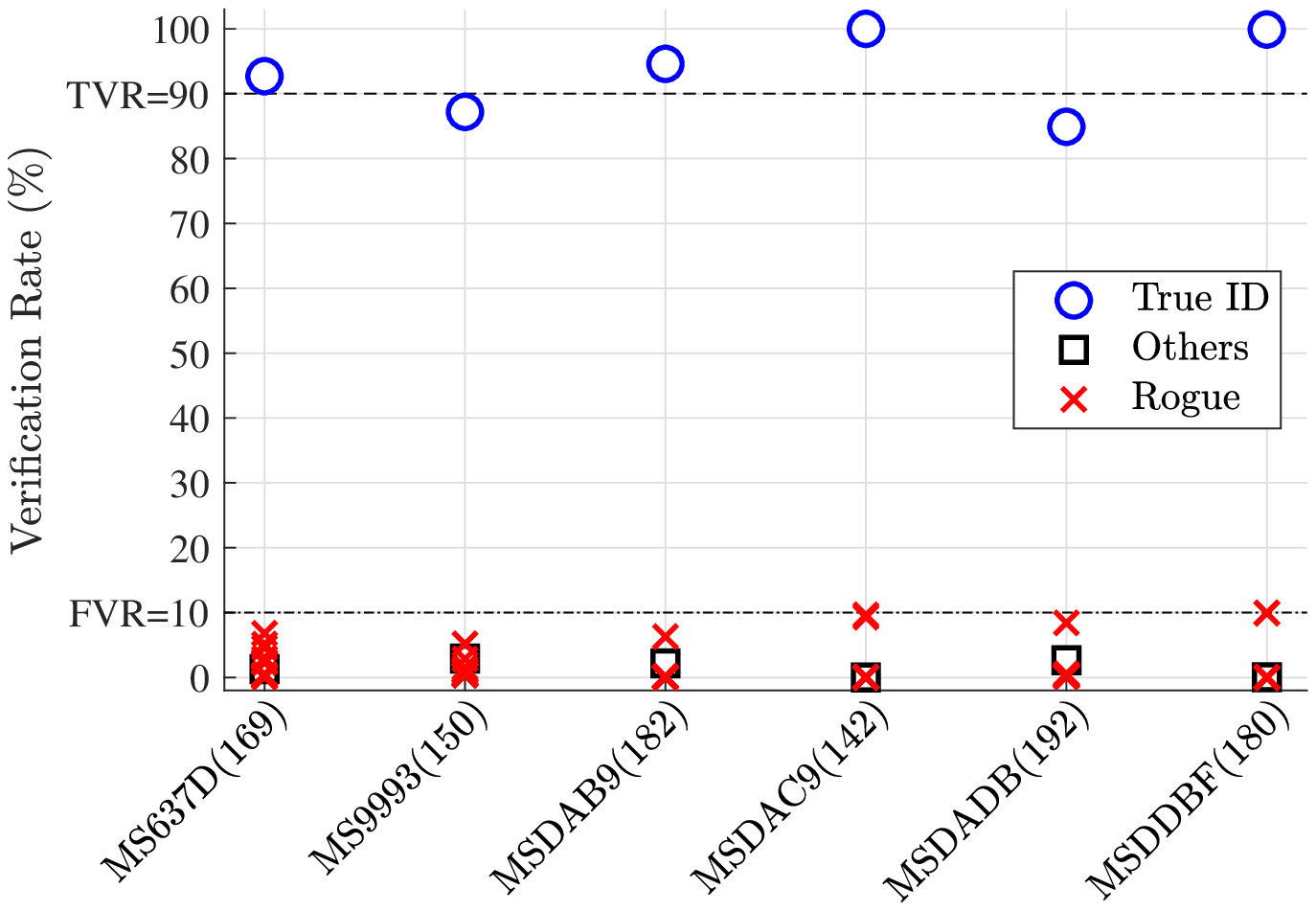}}
\end{subfigure}\hfil 
\begin{subfigure}[Trial \#3 at SNR$=$3~{dB}.]{\label{fig:reliefF_3dB_c}
  \includegraphics[width=0.3\linewidth]{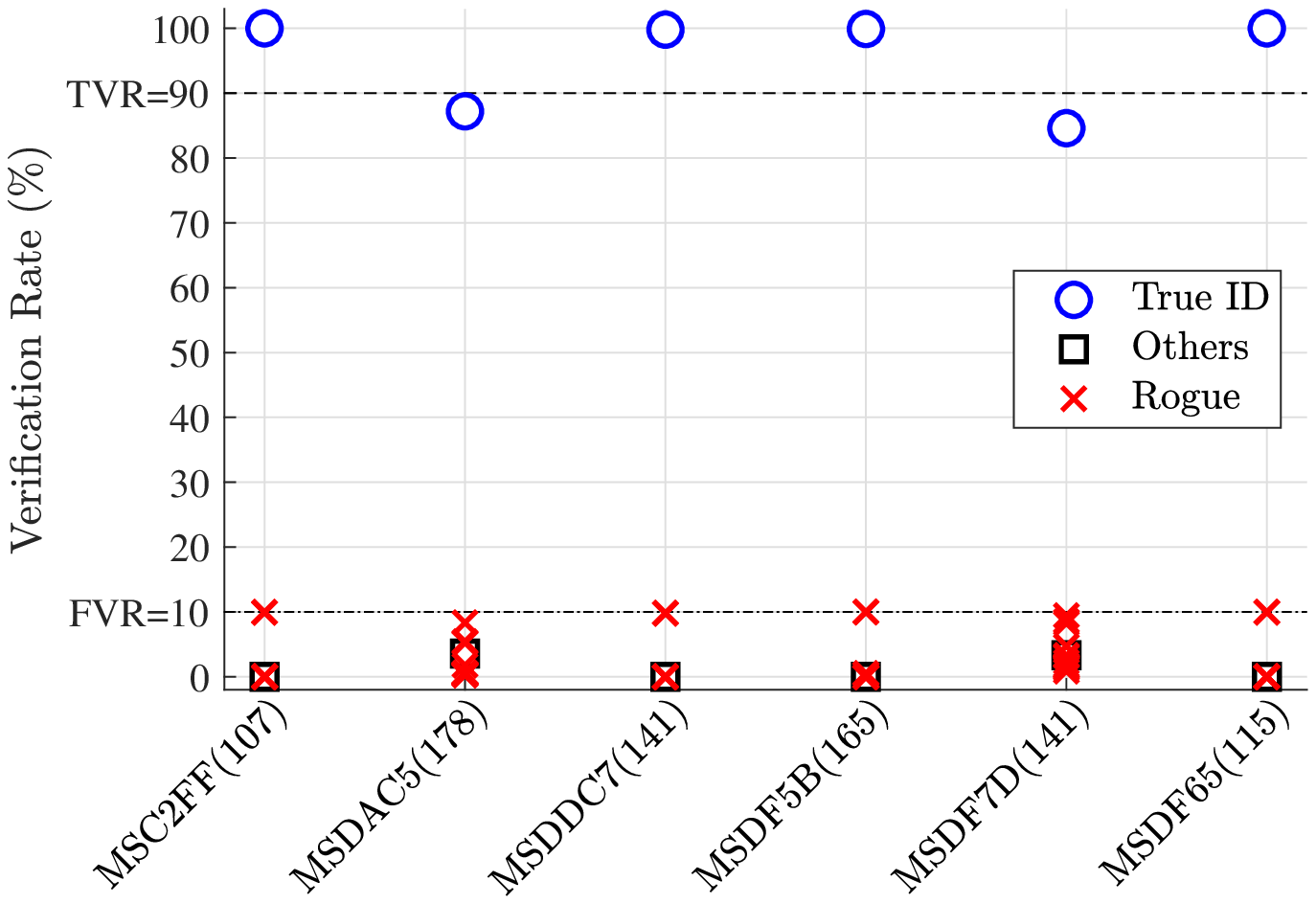}}
\end{subfigure}
\caption{ID verification $({\color{blue}{\circ}})$ and rogue rejection $({\color{red}{\times}})$ performance for the six authorized radios of \emph{all} three trials using Relief-F feature selection at \underline{SNR$=$[3, 6]~{dB}}. The x-axis labels indicate the \emph{claimed} digital ID with the number of retained features in parentheses and ``Others $(\square)$'' indicates the five authorized radios whose ID is not being verified.}
\vspace{-5mm}
\label{fig:reliefF_6dB}
\end{centering}
\end{figure*}
\indent In comparison to the results in \cite{Reising_InfoSec_2015}, the DRA results shown in Fig.~\ref{fig:all_feat_methods} prove superior in that the ID of all six Trial \#1 authorized radios is verified at a TVR$\geq$90\%. A total of 3 and 4 rogue radio attacks fail to be rejected at the FVR$\leq$10\% benchmark in Fig.~\ref{fig:all_feat_methods} and \cite{Reising_InfoSec_2015}, respectively. The worst case DRA-based rogue rejection performance in Fig.~\ref{fig:all_feat_methods} is 87\% versus $\sim$50\% in \cite{Reising_InfoSec_2015}.
\subsection{Evaluation of ID verification Under Degrading SNR}{\label{sec:degrading_snr}}%
Based upon the results in Fig.~\ref{fig:all_feat_methods}, ID verification and rogue rejection performance is assessed using $N_{r}$-dimensional RF-DNA fingerprints whose top ranked features are selected using the: NCA, POEACC, $t$-test, and Relief-F feature selection methods under degrading SNR for all three trials in Table~\ref{tbl:WiMAX_Trials}. The set of all $N_{f}$-dimensional RF-DNA fingerprints contains SNR$\in$[-3, 27]~{dB} in 3~{dB} steps; thus, the degrading SNR assessment is conducted in 3~{dB} steps starting with 18~{dB}. Initially, only the authorized radios of Trial \#1 are used and additional trials are assessed as long as the required TVR and FVR benchmarks are satisfied for the previous trial(s). At the selected SNR, a feature selection method is removed from consideration if either the TVR$\geq$90\% is not met for any of the authorized radios being verified or the FVR$\leq$10\% is not satisfied for the other authorized and rogue radios for any one of the three trials. For SNR$\in$[12, 18]~{dB}, all four of the chosen feature selection approaches met the TVR and FVR benchmarks for each trials' set of authorized and rogue radios. \\
\indent However, at SNR$=$9~{dB}, only POEACC and Relief-F selected $N_{r}$ RF-DNA fingerprint features achieve the TVR$\geq$90\% and FVR$\leq$10\% benchmarks for all three authorized radio trials, as shown in Fig.~\ref{fig:poeacc_reliefF_9dB}. The x-axis indicates the \emph{claimed} digital ID with the number of retained features in parentheses. The SNR$=$9~{dB} results are presented for two reasons: (i) many digital communication standards require a received signal SNR$\geq$10~dB for reliable demodulation performance, e.g., IEEE 802.11 Wi-Fi \cite{802.11}, and (ii) the results presented in \cite{KandahiThings2019} are at SNR$=$9~{dB}, which permits comparative assessment. The work in \cite{KandahiThings2019} only assessed ID verification and rogue rejection performance for Trial \#1 authorized radios. In \cite{KandahiThings2019}, 4 of the 6 authorized radios have their ID verified at a TVR$\geq$90\%. Rogue radio attacks are successfully rejected at FVR$\leq$10\% when spoofing 5 of the 6 authorized radios' digital IDs. When spoofing the ID of MSD905, the 12 rogue radios are rejected at an FVR$=$88\%, which means that the SVM struggles to differentiate a rogue radio from MSD905 when using the DRA selected RF-DNA fingerprint features in \cite{KandahiThings2019}. \\
\indent Due the TVR and FVR performance results presented in Fig.~\ref{fig:poeacc_reliefF_9dB}, ID verification and rogue rejection performance is continued to be assessed for SNR$=$[3, 6]~{dB}. However, for these SNR values the POEACC selected $N_{r}$-dimensional RF-DNA fingerprints no longer satisfied the selected TVR$\geq$90\% and FVR$\leq$10\% benchmarks. Thus, ID verification and rogue radio rejection results are only presented for $N_{r}$-dimensional RF-DNA fingerprints whose features are selected using the Relief-F technique, which are shown in Fig.~\ref{fig:reliefF_6dB}.\\
\indent The SNR$=$6~{dB} results are presented in Fig.~\ref{fig:reliefF_6dB_a}, Fig.~\ref{fig:reliefF_6dB_b}, and Fig.~\ref{fig:reliefF_6dB_c} for Trial \#1, Trial \#2, and Trial \#3, respectively. For the SNR$=$6~{dB} case, the Relief-F selected $N_{r}$ RF-DNA fingerprint features achieve the required TVR$\geq$90\% and FVR$\leq$10\% benchmarks for all authorized and rogue radios across all three of the trials in Table~\ref{tbl:WiMAX_Trials}. The lowest TVR is for MSDF7D of Trial \#3 (Fig.~\ref{fig:reliefF_6dB_c}) with a rate of 93.2\%. All other authorized radios have their IDs verified at TVR$>$93.2\%. For each trial, a total of 72 total rogue radio spoofing attacks occur, which are all successfully defeated at a FVR$\leq$10\%. \\
\indent The SNR$=$3~{dB} results are shown for: Trial \#1 in Fig.~\ref{fig:reliefF_3dB_a}, Trial \#2 in Fig.~\ref{fig:reliefF_3dB_b}, and Trial \#3 in Fig.~\ref{fig:reliefF_3dB_c}. For Trial \#1 the ID of all six authorized radios is correctly verified at a TVR$\geq$90\%. The lowest true verification performance occurring for radio MS66E7 at a TVR$=$91\%. All rogue radios, representing 12 spoofing attacks per authorized radio for a total of 72, are successfully rejected at a FVR$\leq$10\% at SNR$=$3~{dB}. Trial \#1's worst case rejection performance is at an FVR$=$7.4\%, which occurs when a rogue radio spoofs the ID of the authorized radio MS6387. In terms of Trial \#2 presented in Fig.~\ref{fig:reliefF_3dB_b}, the IDs of four out of the six authorized radios is verified at a TVR$\geq$90\%. The two remaining radios, MS9993 and MSDADB, are verified at rates of 87.2\% and 84.9\%, respectively. Despite these two authorized radios not being verified at or above the selected TVR benchmark, all rogue radios are still successfully rejected within the FVR$\leq$10\% benchmark or better. Worst case rogue rejection performance is an FVR$=$9.93\% when the ID of authorized radio MSDDBF is spoofed by one of the 12 rogue radios. For Trial \#3 in Fig.~\ref{fig:reliefF_3dB_c}, the TVR$\geq$90\% benchmark is achieved when verifying the ID of four out of the six authorized radios. Authorized radios MSDAC5 and MSDF7D are verified at TVR$=$87.2\% and TVR$=$84.6\%, respectively. Similar to Trial \#2, all of the rogue radio attacks on Trial \#3 authorized radios are successfully rejected at the desired FVR$\leq$10\% or lower. Worst case rogue rejection performance is an FVR$=$10\% when the IDs of authorized radios: MSC2FF, MSDDC7, MSDF5B, and MSDF65 are spoofed by a rogue radio. When considering the results for Trial \#2 and Trial \#3, with respect to the threat model presented in Sect.~\ref{sec:threat_model}, the ability to successfully reject all of the rogue radio attacks, at the selected FVR benchmark, is of greater import than verifying the IDs of all authorized radios at the selected TVR$\geq$90\% benchmark. The success of the Releif-F selected RF-DNA fingerprint features is attributed to the algorithm's use of the $k$ nearest in-class and out-of-class neighbors  when determining the feature weights; thus, making making it more robust under degrading SNR conditions.
\section{Conclusion}\label{sec:conclusion}%
Over the next five years the number of deployed IoT devices is estimated to reach 75 billion. The vast majority of these and current IoT devices lack robust security to protect them, and the associated networks, from exploitation by nefarious actors. This work presents a PHY layer IoT security approach capable of defeating digital ID spoofing attacks through the use of feature reduced RF-DNA fingerprints and an SVM classifier. A total of eight feature selection approaches are assessed to determine the RF-DNA fingerprint features that facilitate authorized radio ID verification at a TVR$\geq$90\% while simultaneously rejecting all rogue radio digital ID spoofing attacks at a FVR$\leq$10\% at the lowest SNR possible. Unlike prior works that use RF fingerprints formed using a fixed number and/or composition of features for all authorized radios, this work allows the number and composition of selected RF fingerprint features to be driven by individual authorized radio ID verification performance. Selection is driven by the distribution of the margin values generated from the SVM model and the authorized radios' $N_{r}$-dimensional RF-DNA fingerprints, i.e., without knowledge of rogue radios' RF-DNA fingerprints. This work successfully demonstrates 100\%: (i) correct ID verification, i.e., TVR$\geq$90\%, across three trials of six randomly selected authorized radios at SNR$\geq$6dB, and (ii) rejection, i.e., FVR$\leq$10\%, of 72 rogue radio ID spoofing attacks per authorized radio SNR$\geq$3dB using $N_{r}$-dimensional RF-DNA fingerprints whose features are selected using the Relief-F algorithm. Such performance is unseen in previous RF fingerprint-based ID verification publications. Future work will focus on ID verification and rogue radio rejection for networks comprised of: (i) differing numbers of authorized radios, and (ii) radios that are heterogeneous in manufacturer and/or model. 

\balance
\bibliographystyle{IEEEtran}
\bibliography{references_v02}\label{sec:references_v02}

\begin{thebibliography}{10}
\providecommand{\url}[1]{#1}
\csname url@samestyle\endcsname
\providecommand{\newblock}{\relax}
\providecommand{\bibinfo}[2]{#2}
\providecommand{\BIBentrySTDinterwordspacing}{\spaceskip=0pt\relax}
\providecommand{\BIBentryALTinterwordstretchfactor}{4}
\providecommand{\BIBentryALTinterwordspacing}{\spaceskip=\fontdimen2\font plus
\BIBentryALTinterwordstretchfactor\fontdimen3\font minus
  \fontdimen4\font\relax}
\providecommand{\BIBforeignlanguage}[2]{{%
\expandafter\ifx\csname l@#1\endcsname\relax
\typeout{** WARNING: IEEEtran.bst: No hyphenation pattern has been}%
\typeout{** loaded for the language `#1'. Using the pattern for}%
\typeout{** the default language instead.}%
\else
\language=\csname l@#1\endcsname
\fi
#2}}
\providecommand{\BIBdecl}{\relax}
\BIBdecl

\bibitem{DoD_IoT_2016}
{Chief Information Officer, U.S. Department of Defense}, ``{DoD Policy
  Recommendations for The Internet of Things (IoT)},''
  https://www.hsdl.org/?view\&did=799676, Dec. 2016.

\bibitem{Gartner_2015}
{Gartner Research}, ``{Gartner Says 6.4 Billion Connected``Things'' Will Be in
  Use in 2016, Up 30 Percent From 2015},'' Nov. 2015.

\bibitem{Juniper_2016}
{Juniper Research}, ``{`Internet of Things' Connected Devices to Triple by
  2021, Reaching Over 46 Billion Units},'' {Dec.} {2016}.

\bibitem{Statista_IoT_2019}
{Statista}, ``{Internet of Things (IoT) connected devices installed base
  worldwide from 2015 to 2025 (in billions)},''
  https://www.statista.com/statistics/471264/iot-number-of-connected-devices-worldwide/,
  2019.

\bibitem{Rawlinson_2014}
{Rawlinson, K.}, ``{HP Study Reveals 70 Percent of Internet of Things Devices
  Vulnerable to Attack},'' {Jul.} 2014.

\bibitem{Ray_CIC_2019}
{Ray, I., D. Kar, J. Peterson, and S. Goeringer}, ``{Device Identity and Trust
  in IoT-sphere Forsaking Cryptography},'' in \emph{International Conference on
  Collaboration and Internet Computing (CIC)}, 2019.

\bibitem{Larsen_CNN_2017}
{Larsen, S.}, ``{A smart fish tank left a casino vulnerable to hackers},''
  Website:
  https://money.cnn.com/2017/07/19/technology/fish-tank-hack-darktrace/index.html,
  Jul 2017.

\bibitem{Wright_book_2015}
{Wright, J., and J. Cache,}, \emph{{Hacking Wireless Exposed: Wireless Security
  Secrets \& Solutions}}, 3rd~ed.\hskip 1em plus 0.5em minus 0.4em\relax
  {McGraw-Hill}, 2015.

\bibitem{Stanislav_Rapid_2015}
{Stanislav, M., and T. Beardsley}, ``{Hacking IoT: A Case Study on Baby
  Monitoring Exposures and Vulnerabilities}.''

\bibitem{Wright_Killer_2019}
\BIBentryALTinterwordspacing
{Wright, J.}, ``{KillerBee: Practical ZigBee Exploitation Framework or
  'Wireless Hacking and the Kinetic World'}.'' [Online]. Available:
  \url{https://www.inguardians.com/works/}
\BIBentrySTDinterwordspacing

\bibitem{Simon_2016}
{Simon, S.}, ``{'Internet Of Things' Hacking Attack Led To Widespread Outage Of
  Popular Websites},'' Oct 2016. [Online]. Available:
  https://www.wbur.org/npr/498954197/internet-outage-update-internet-of-things-hacking-attack-led-to-outage-of-popular.

\bibitem{Shipley_DEFCON_2014}
\BIBentryALTinterwordspacing
{Shipley, P.}, ``{Insteon: False Security and Deceptive Documentation},''
  \emph{DEFCON 23}, 2014. [Online]. Available:
  \url{https://www.youtube.com/watch?v=dy1LTQLmPtM}
\BIBentrySTDinterwordspacing

\bibitem{Shipley_GitHub}
\BIBentryALTinterwordspacing
------, ``{Tools for Insteon RF},'' 2015. [Online]. Available:
  \url{https://github.com/evilpete/insteonrf}
\BIBentrySTDinterwordspacing

\bibitem{Krebs_Mirai_2017}
\BIBentryALTinterwordspacing
{Krebs, B.}, ``{Mirai IoT Botnet Co-Authors Plead Guilty},'' 2017. [Online].
  Available:
  \url{https://krebsonsecurity.com/2017/12/mirai-iot-botnet-co-authors-plead-guilty/}
\BIBentrySTDinterwordspacing

\bibitem{Talbot_CandS_2017}
{Talbot, C., M. Temple, T. Carbino, and J. Betances}, ``{Detecting Rogue
  Attacks on Commercial Wireless Insteon Home Automation Systems},''
  \emph{Computers \& Security}, vol.~74, 10 2017.

\bibitem{Sa_Access_2019}
{Sa, K., D. Lang, C. Wang, and Y. Bai}, ``{Specific Emitter Identification
  Techniques for the Internet of Things},'' \emph{IEEE Access}, 2019.

\bibitem{ToonsKins95}
{Toonstra J. and W. Kinsnew}, ``{Transient Analysis and Genetic Algorithms for
  Classification},'' in \emph{IEEE Conf on Communications, Power \& Computing},
  May 1995.

\bibitem{UretenIEE99}
{Ureten O. and N. Serinken}, ``{Detection of Radio Transmitter Turn-On
  Transients},'' \emph{IEE Electronics Letters}, vol.~35, no.~23, Nov 1999.

\bibitem{DudczykSEI2}
{Dudczyk J., J. Matuszewski and M. Wnuk}, ``{Applying the Radiated Emission to
  the Specific Emitter Identification}.''\hskip 1em plus 0.5em minus
  0.4em\relax Int'l Conf on Microwaves, Radar \& Wireless Communications, May
  2004.

\bibitem{Jeffery_MobiCom_2007}
{Jeffrey, P., G. Ben, G. Ramakrishna, S. Srinivasan and W. David}, ``{802.11
  User Fingerprinting},'' in \emph{{ACM Int'l Conf on Mobile Computing \&
  Networking}}, Jun 2013.

\bibitem{JanaMobi08}
{Jana, S. and S. Kasera}, ``{On Fast and Accurate Detection of Unauthorized
  Wireless Access Points Using Clock Skews},'' in \emph{ACM Int'l Conf on
  Mobile Computing \& Networking}, Sep 2008.

\bibitem{BrikMobi08}
{Brik, V., S. Banerjee, M. Gruteserand S. Oh}, ``{Wireless Device
  Identification with Radiometric Signatures},'' in \emph{ACM Int'l Conf on
  Mobile Computing \& Networking}, Sep 2008.

\bibitem{Suski_IJESDF_2008}
{Suski W. II, M. Temple, M. Mendenhall and R. Mills}, ``{RF Fingerprinting
  Commercial Communication Devices to Enhance Electronic Security},''
  \emph{Int'l J. Electronic Security \& Digital Forensics}, {vol. 1, no. 3,
  2008}.

\bibitem{DanevIPSN09}
{Danev B. and S. Kapkun}, ``{Transient-Based Identification of Wireless Sensor
  Nodes},'' in \emph{ACM Int'l Conf on Info Processing in Sensor Networks}, Apr
  2009.

\bibitem{Klein_ICC_2009}
{Klein R., M. Temple, M. Mendenfhall and D. Reising}, ``Sensitivity {A}nalysis
  of {B}urst {D}etection and {RF} {F}ingerprinting {C}lassification
  {P}erformance,'' in \emph{IEEE Int'l Conf on Communications}, Jun 2009.

\bibitem{Liu_SEI_2009}
{Liu, M. and J. Doherty}, ``{Nonlinearity Estimation for Specific Emitter
  Identification in Multipath Environment},'' in \emph{IEEE Sarnoff Symp}, Mar
  2009.

\bibitem{Liu_SEI_2011}
------, ``{Nonlinearity Estimation for Specific Emitter Identification in
  Multipath Channels},'' \emph{IEEE Trans on Info Forensics \& Security},
  vol.~6, no.~3, Sep 2011.

\bibitem{Kennedy_2010}
{Kennedy, I. and A. Kuzminskiy}, ``{RF Fingerprint Detection in a Wireless
  Multipath Channel},'' in \emph{Int'l Symp on Wireless Comm Systems}, Sep
  2010.

\bibitem{Reising_IJESDF_2010}
{Reising, D., M. Temple,and M. Mendenhall}, ``{Improved wireless security for
  GMSK-based devices using RF fingerprinting},'' \emph{Int. J. Electron. Secur.
  Digit. Forensic}, vol.~3, no.~1, 2010.

\bibitem{Reising_Dissertation}
{Reising, D.}, ``{Exploitation of RF-DNA for Device Classification and
  Verification Using GRLVQI Processing},'' Ph.D. dissertation, {Air Force
  Institute of Technology}, {Dec. 2012}.

\bibitem{Williams_NSS_2010}
{Williams M., S. Munns, M. Temple and M. Mendenhall}, ``{RF-DNA Fingerprinting
  for Airport WiMax Communications Security},'' in \emph{Int'l Conf on Net \&
  Sys Security}, Sep 2010.

\bibitem{Takahashi_CompApps_2010}
{Takahashi, D., Y. Xiaoa, Y. Zhang, P. Chatzimisios, and H. Chend}, ``{IEEE
  802.11 User Fingerprinting and its Applications for Intrusion Detection},''
  \emph{{Computers \& Math with Applications}}, {vol. 60, no. 2, 2010}.

\bibitem{TekbasIEE}
{Tekbas, O., O. Ureten and N. Serinken}, ``{Improvement of Transmitter
  Identification System for Low SNR Transients},'' \emph{IEE Electronics
  Letters}, {vol. 40, no. 3, Jul 2004}.

\bibitem{Ellis_RadioSci}
{Ellis K. and N. Serinken}, ``{Characteristics of Radio Transmitter
  Fingerprints},'' \emph{Radio Science}, {vol. 36, no. 4, 2001}.

\bibitem{Soli_IEEE}
{Soliman, S. and S-Z. Hsue}, ``{Signal Classification Using Statistical
  Moments},'' \emph{IEEE Trans on Communications}, {vol. 40, no. 5, May 1992}.

\bibitem{CanadaSEI}
{Defence R\&D Canada - Ottawa}, ``{Interferometric Intrapulse Radar Receiver
  for Specific Emitter Identification and Direction-Finding},'' \emph{{Fact
  Sheet REW 224}}, Jun 2007.

\bibitem{azzouz}
{Azzouz E. and A. Nandi}, \emph{{Automatic Modulation Recognition of
  Communication Signals}}.\hskip 1em plus 0.5em minus 0.4em\relax Kluwer
  Academic Publishers, 1996.

\bibitem{Wheeler_ICNC_2017}
{Wheeler C. and D. Reising}, ``{Assessment of the impact of CFO on RF-DNA
  fingerprint classification performance},'' in \emph{Int'l Conf on Computing,
  Networking \& Communications}, Jan 2017.

\bibitem{JafariMILCOM2018}
{Jafari, H., O. Omotere, D. Adesina, H-H. Wu, and L. Qian}, ``{IoT Devices
  Fingerprinting using Deep Learning},'' in \emph{IEEE Military Comm Conf
  (MILCOM)}, Oct 2018.

\bibitem{Pan_2019}
{Pan, Y., S. Yang, H. Peng, T. Li and W. Wang}, ``{Specific Emitter
  Identification Based on Deep Residual Networks},'' \emph{IEEE Access},
  vol.~7, 2019.

\bibitem{KoseAccess2019}
{Köse, M., S. Taşcioğlu and Z. Telatar}, ``{RF Fingerprinting of IoT Devices
  Based on Transient Energy Spectrum},'' \emph{IEEE Access}, vol.~7, 2019.

\bibitem{Fadul_WCNC_2019}
{Fadul M., D. Reising, D. Loveless and A. Ofoli}, ``{RF-DNA Fingerprint
  Classification of OFDM Signals Using a Rayleigh Fading Channel Model},'' in
  \emph{IEEE Wireless Communications and Networking Conf (WCNC)}, April 2019.

\bibitem{Tian_IOTJournal_2019}
{Tian, Q., Y. Lin, X. Guo, J. Wen, Y. Fang, J. Rodriguez, and S. Mumtaz},
  ``{New Security Mechanisms of High-Reliability IoT Communication Based on
  Radio Frequency Fingerprint},'' \emph{IEEE IoT Journal}, 2019.

\bibitem{Wilson_GLOBECOM_2019}
{Wilson, A., D. Reising, and T. Loveless}, ``{Integration of Matched Filtering
  within the RF-DNA Fingerprinting Process},'' in \emph{IEEE Global
  Telecommunications Conf (GLOBECOM)}, ACCEPTED - Jul. 2019.

\bibitem{Kroon_AICompSci_2010}
{Kroon, B., S. Bergin, I. Kennedy, and G. O'Mahony Zamora}, ``{Steady State RF
  Fingerprinting for Identity Verification: One Class Classifier versus
  Customized Ensemble},'' in \emph{A.I. \& Cognitive Science}, 2010.

\bibitem{Cobb_IFS_11}
{Cobb W., E. Laspe, R. Baldwin, M. Temple and Y. Kim}, ``{Intrinsic Physical
  Layer Authentication of ICs},'' \emph{IEEE Trans on Information Forensics and
  Security}, vol.~2, no.~4, p.~7, Dec 2011.

\bibitem{Dubendorfer_MILCOM_2012}
{Dubendorfer C., B. Ramsey and M. Temple}, ``{An RF-DNA Verification Process
  for ZigBee Networks},'' in \emph{Proc of 2012 IEEE Military Comm Conf
  (MILCOM12)}, Oct 2012.

\bibitem{Rehman_JoCompSysSci_2014}
{Rehman, S., K. Sowerby, and C. Coghill}, ``{Analysis of impersonation attacks
  on systems using RF fingerprinting and low-end receivers},'' \emph{Journal of
  Computer and System Sciences}, vol.~80, p. 591–601, 05 2014.

\bibitem{Reising_InfoSec_2015}
{Reising D., M. Temple and J. Jackson}, ``{Authorized and Rogue Device
  Discrimination Using Dimensionally Reduced RF-DNA Fingerprints},''
  \emph{{IEEE Trans on Info Forensics \& Security}}, {vol. 10, no. 6, 2015}.

\bibitem{Patel_TransOnReli_2015}
{Patel, H., M. Temple, and R. Baldwin}, ``{Improving ZigBee Device Network
  Authentication Using Ensemble Decision Tree Classifiers With Radio Frequency
  Distinct Native Attribute Fingerprinting},'' \emph{{IEEE Trans on
  Reliability}}, vol.~64, no.~1, March 2015.

\bibitem{Baldini_Sciences_2018}
{Baldini, G., R. Giuliani,and G. Steri}, ``{Physical Layer Authentication and
  Identification of Wireless Devices Using the Synchrosqueezing Transform},''
  \emph{Applied Sciences}, vol.~8, p. 2167, 11 2018.

\bibitem{Merchant_JSTSP_2018}
{Merchant, K., S. Revay, G. Stantchev, and B. Nousain}, ``{Deep Learning for RF
  Device Fingerprinting in Cognitive Communication Networks},'' \emph{IEEE J.
  of Selected Topics in Signal Processing}, vol.~12, no.~1, Feb 2018.

\bibitem{AndrewsWiSec2019}
{Andrews, S., R. Gerdes, and M. Li}, ``{Crowdsourced Measurements for Device
  Fingerprinting},'' in \emph{{ACM Conf on Security and Privacy in Wireless and
  Mobile Network (WiSec)}}, May 2019.

\bibitem{KandahiThings2019}
{Kandah, F., J. Cancelleri, D. Reising, A. Altarawneh, and A. Skjellum}, ``{A
  Hardware-Software Co-design Approach to Identity, Trust, and Resilience for
  IoT/CPS at Scale},'' in \emph{International Conference on Internet of Things
  (iThings)}, July 2019.

\bibitem{Wang_CommsMag_2016}
{Wang, X., P. Hao, and L. Hanzo}, ``{Physical-Layer Authentication for Wireless
  Security Enhancement: Current Challenges and Future Developments},''
  \emph{IEEE Communications Magazine}, vol.~54, Jun 2016.

\bibitem{Xu_CommsTuts_2016}
{Xu, Q., R. Zheng, W. Saad, and Z. Han}, ``{Device Fingerprinting in Wireless
  Networks: Challenges and Opportunities},'' \emph{IEEE Communications Surveys
  \& Tutorials}, vol.~18, no.~1, 2016.

\bibitem{Clancy_CrownCom_2008}
{Clancy, T. C. and N. Goergen}, ``{Security in Cognitive Radio Networks:
  Threats and Mitigation},'' in \emph{International Conference on Cognitive
  Radio Oriented Wireless Networks and Communications (CrownCom)}, 2008.

\bibitem{Xie_TransMobi_2020}
{Xie, T., G. Tu, C. Li, and C. Peng}, ``{How Can IoT Services Pose New Security
  Threats In Operational Cellular Networks?}'' \emph{IEEE Transactions on
  Mobile Computing}, 2020.

\bibitem{Tian_Sensors_2020}
{Tian, Q., Y. Lin, X. Guo, J. Wang, O. AlFarraj, and A. Tolba}, ``{An Identity
  Authentication Method of a MIoT Device Based on Radio Frequency Fingerprint
  Technology},'' \emph{Sensors}, vol.~20, no.~4, Feb 2020.

\bibitem{Agilent}
``{Agilent E3238 Signal Intercept and Collection Solutions},'' 2004.

\bibitem{Bastiaans96}
{Bastiaans, M. J.}, ``{Discrete Gabor Transform and Discrete Zak Transform},''
  in \emph{IEEE Int'l Conf on Signal \& Image Proc Applications}, 1996.

\bibitem{dhsPC}
{Duda R., P. Hart and D. Stork}, \emph{Pattern Classification}, 2nd~ed.\hskip
  1em plus 0.5em minus 0.4em\relax John Wiley \verb"&" Sons, Inc., 2001.

\bibitem{YangJCP2012}
{Yang, W., K. Wang, and W. Zuo}, ``{Neighborhood Component Feature Selection
  for High-Dimensional Data},'' \emph{{Journal of Computers (JCP)}}, vol.~7,
  2012.

\bibitem{MucciardiITC}
{Mucciardi, A. N. and E. E. Gose}, ``{A Comparison of Seven Techniques for
  Choosing Subsets of Pattern Recognition Properties}.''\hskip 1em plus 0.5em
  minus 0.4em\relax IEEE Trans on Computers, 1971.

\bibitem{Comaniciu_CVPR_2000}
{Comaniciu, D., V. Ramesh, and P. Meer}, ``{Real-Time Tracking of Non-Rigid
  Objects using Mean Shift},'' in \emph{Proceedings IEEE Conference on Computer
  Vision and Pattern Recognition (CVPR)}, vol.~2, June 2000.

\bibitem{Derrick_TQMP_2016}
{Derrick, B. and P. White}, ``{Why Welch's Test is Type I Error Robust},''
  \emph{The Quantitative Methods for Psychology (TQMP)}, vol.~12, no.~1, 2016.

\bibitem{Allwood_APStats_2008}
\BIBentryALTinterwordspacing
{Allwood, M.}, ``{The Satterthwaite Formula for Degrees of Freedom in the
  Two-Sample $t$-Test},'' \emph{AP Statistics}, 2008. [Online]. Available:
  \url{http://apcentral.collegeboard.com/apc/public/repository/ap05_stats_allwood_fin4prod.pdf}
\BIBentrySTDinterwordspacing

\bibitem{Kononenko}
{Kononenko, I.}, ``{Estimating Attributes: Analysis and Extensions of
  Relief}.''\hskip 1em plus 0.5em minus 0.4em\relax European Conference on
  Machine Learning, 1994.

\bibitem{Hastie}
{Hastie T., R. Tibshirani and J. Friedman}, \emph{{The Elements of Statistical
  Learning; Data Mining, Inference, and Prediction}}.\hskip 1em plus 0.5em
  minus 0.4em\relax {Springer-Verlag, New York, New York, USA}, 2001.

\bibitem{watkins}
{Watkins D. S.}, \emph{{Fundamentals of Matrix Computations}}, 2nd~ed.\hskip
  1em plus 0.5em minus 0.4em\relax John Wiley \& Sons, Inc., 2002.

\bibitem{Yang_JCP_2012}
{Yang, W., K. Wang, and W. Zuo}, ``{Neighborhood Component Feature Selection
  for High-Dimensional Data},'' \emph{J. of Computers}, vol.~7, 2012.

\bibitem{GonzalezThesis}
{Gonz\'{a}lez J. A.}, ``{Numerical Analysis for Relevant Features in Intrusion
  Detection (NARFid)},'' Master's thesis, Air Force Institute of Technology,
  2950 Hobson Way, WPAFB, OH, March 2009.

\bibitem{Welch_Biometrika_1947}
\BIBentryALTinterwordspacing
{Welch, B.}, ``{The Generalization of `Student's' Problem when Several
  Different Population Variances are Involved},'' \emph{Biometrika}, vol.~34,
  no. 1/2, 1947. [Online]. Available: \url{http://www.jstor.org/stable/2332510}
\BIBentrySTDinterwordspacing

\bibitem{Ruxton_BE_2006}
{Ruxton, G.}, ``{The Unequal Variance $t$-test is an Underused Alternative to
  Student's $t$-test and the Mann–Whitney $U$ test},'' \emph{Behavioral
  Ecology}, vol.~17, no.~4, May 2006.

\bibitem{Kira01}
{Kira, K. and L. A. Rendell}, ``{The Feature Selection Problem: Traditional
  Methods and a New Algorithm}.''\hskip 1em plus 0.5em minus 0.4em\relax
  Proceedings of the Tenth National Conference on Artificial Intelligence AAAI,
  1992.

\bibitem{Kira02}
------, ``{A Practical Approach to Feature Selection}.''\hskip 1em plus 0.5em
  minus 0.4em\relax Assorted Conferences and Workshops, 1992.

\bibitem{Durgabai_IJARCCE_2014}
{Durgabai, R. and Y. RaviBhushan}, ``{Feature Selection using ReliefF
  Algorithm},'' vol.~3, no.~10, Oct 2014.

\bibitem{Stief_MMAR_2018}
{Stief, A., J. Ottewill, and J. Baranowski}, ``{Relief F-Based Feature Ranking
  and Feature Selection for Monitoring Induction Motors},'' in
  \emph{International Conference on Methods Models in Automation Robotics
  (MMAR)}, Aug 2018.

\bibitem{Christianini}
{Christianini, N., and J. Shawe-Taylor}, \emph{{An Introduction to Support
  Vector Machines and Other Kernel-Based Learning Methods}}.\hskip 1em plus
  0.5em minus 0.4em\relax Cambridge, UK: Cambridge University Press, 2000.

\bibitem{Tharwat_AIComms_2017}
{Tharwat, A., T. Gaber, A. Ibrahim, and A. Hassanien}, ``{Linear discriminant
  analysis: A detailed tutorial},'' \emph{{A.I. Comms}}, vol.~30, May 2017.

\bibitem{802.11}
\emph{IEEE Std 802.11-2007, Local and Metropolitan Area Networks, Part 11:
  Wireless LAN Medium Access Control (MAC) and Physical Layer (PHY)
  Specifications}, IEEE, Jun 2007.

\end{thebibliography}

\end{document}